\documentclass[]{aaV6.1}
\usepackage{graphicx}
\newcommand{\kmprs}  {\mbox{\rm km\,s$^{-1}$}}
\newcommand{\micro}{$\xi_{\rm turb}$}

\newcommand{\feh} {\mbox{\rm [Fe/H]}}

\newcommand{\logS}{log\,$\epsilon ({\rm S})$}
\newcommand{\logSlte}{log\,$\epsilon ({\rm S})_{\rm LTE}$}
\newcommand{\logZn}{log\,$\epsilon ({\rm Zn})$}

\newcommand{\logFesun}{log\,$\epsilon ({\rm Fe})_{\odot}$}
\newcommand{\logSsun}{log\,$\epsilon ({\rm S})_{\odot}$}
\newcommand{\logZnsun}{log\,$\epsilon ({\rm Zn})_{\odot}$}
\newcommand{\znh} {\mbox{\rm [Zn/H]}}

\newcommand{\sfe} {\mbox{\rm [S/Fe]}}

\newcommand{\ofe} {\mbox{\rm [O/Fe]}}
\newcommand{\szn} {\mbox{\rm [S/Zn]}}

\newcommand{\znfe} {\mbox{\rm [Zn/Fe]}}

\newcommand{\mgfe} {\mbox{\rm [Mg/Fe]}}
\newcommand{\sife} {\mbox{\rm [Si/Fe]}}
\newcommand{\cafe} {\mbox{\rm [Ca/Fe]}}
\newcommand{\tife} {\mbox{\rm [Ti/Fe]}}
\newcommand{\alphafe} {\mbox{\rm [$\alpha$/Fe]}}
\newcommand{\teff}  {\mbox{$T_{\rm eff}$}}
\newcommand{\logteff} {\mbox{${\rm log}\,T_{\rm eff}$}}
\newcommand{\logg}  {\mbox{{\rm log}\,$g$}}
\newcommand{\HI} {\ion{H}{i}}

\newcommand{\SI} {\ion{S}{i}} 
\newcommand{\SII} {\ion{S}{ii}}

\newcommand{\FeI} {\ion{Fe}{i}}
\newcommand{\FeII} {\ion{Fe}{ii}}

\newcommand{\ZnI} {\ion{Zn}{i}}
\newcommand{\ZnII} {\ion{Zn}{ii}}

\newcommand{\Mv} {\mbox{$M_V$}}

\newcommand{\water}{\mbox{\rm H$_2$O}}

\newcommand{\hbeta}{\mbox{\rm{H}$\beta$}}
\newcommand{\pzeta}{\mbox{\rm{P}$\zeta$}}
\newcommand{\VK}{\mbox{($V\!-\!K)$}}
\newcommand{\VKO}{\mbox{($V\!-\!K)_0$}}
\newcommand{\JK}{\mbox{($J\!-\!K)$}}
\newcommand{\by}{\mbox{($b\!-\!y)$}}

\def\ltsima{$\; \buildrel < \over \sim \;$}
\def\simlt{\lower.5ex\hbox{\ltsima}}
\def\gtsima{$\; \buildrel > \over \sim \;$}
\def\simgt{\lower.5ex\hbox{\gtsima}}
\begin{document}

\title{Sulphur and zinc abundances in Galactic halo stars revisited
\thanks{Based on observations collected at the European
Southern Observatory at Paranal, Chile (programmes No. 67.D-0106, 73.D-0024
and CRIRES science verification program 60.A-9072)}
\fnmsep\thanks{Table 1 and Appendices A, B, and C are only available
in electronic form at http://www.aanda.org}
}


\author{P.E.~Nissen \inst{1} \and C.~Akerman \inst{2}
\and M.~Asplund \inst{3} \and D.~Fabbian \inst{3} \and F.~Kerber \inst{4}
\and H.U.~K\"{a}ufl \inst{4}  \and M. Pettini \inst{2}}


\institute{
Department of Physics and Astronomy, University of Aarhus, DK--8000
Aarhus C, Denmark. 
\email{pen@phys.au.dk}
\and Institute of Astronomy, University of Cambridge, Madingley Road, Cambridge, CB3 0HA, UK.
\and
Research School of Astronomy and Astrophysics,
Australian National University, 
Cotter Road, Weston, ACT 2611, Australia.
\and
European Southern Observatory, Karl-Schwarzschild-Str. 2, D-85748 Garching,
Germany.}

\date{Received 23 February 2007 / Accepted 2 April 2007}
\abstract
{}
{Based on a new set of sulphur abundances in very metal-poor stars and
an improved analysis of previous data, we aim at resolving current
discrepancies on the trend of S/Fe vs. Fe/H and thereby gain better 
insight into the nucleosynthesis of sulphur. The trends of Zn/Fe and 
S/Zn will also be studied.}
{High resolution VLT/UVES spectra of 40 main-sequence stars with 
$-3.3 <$ [Fe/H] $< -1.0$
are used to derive S abundances from the weak $\lambda 8694.6$ \SI\ line
and the stronger $\lambda \lambda 9212.9,9237.5$ pair of \SI\ lines.
For one star, the S abundance is also derived from the \SI\ triplet at 
1.046\,$\mu$m recently observed with the VLT infrared echelle 
spectrograph CRIRES. Fe and Zn abundances are derived from lines in the
blue part of the UVES spectra, and effective temperatures are obtained from
the profile of the H$\beta$ line.}
{Comparison of sulphur abundances from the weak and strong \SI\ lines
provides important constraints on non-LTE effects. 
The high sulphur abundances reported by others for some 
metal-poor stars are not confirmed; instead, when taking non-LTE corrections
into account, the Galactic halo stars
distribute around a plateau at [S/Fe]$\sim +0.2$\,dex
with a scatter of 0.07\,dex only. [Zn/Fe] is close to zero for metallicities
in the range $-2.0 <$ [Fe/H] $< -1.0$ but increases to a level of 
[Zn/Fe]$\sim +0.1$ to +0.2\,dex in the range $-2.7 <$ [Fe/H] $< -2.0$. 
At still lower metallicities [Zn/Fe] rises steeply to a value of 
[Zn/Fe]$\sim +0.5$\,dex at [Fe/H] = $-3.2$.}
{The trend of S/Fe vs. Fe/H corresponds to the trends of Mg/Fe, Si/Fe, 
and Ca/Fe and indicates that sulphur in Galactic halo stars has been made by 
$\alpha$-capture processes in massive SNe. The observed scatter in
S/Fe is much smaller than predicted from current
stochastic models of the chemical evolution of the early Galaxy, 
suggesting that either the models or the calculated yields of massive SNe
should be revised.
We also examine the behaviour of S/Zn and find that
departures from the solar ratio are
significantly reduced at all metallicities if non-LTE
corrections to the abundances of these two elements
are adopted. This effect, if confirmed, would reduce the 
usefulness of the S/Zn ratio as a diagnostic of past
star-formation activity, but would bring closer together
the values measured in damped Lyman-alpha systems and
in Galactic stars.}
 

\maketitle

\section{Introduction}
\label{introduction}
Despite several recent papers on sulphur abundances in Galactic
stars, there is still no agreement on the trend of \sfe\ vs. \feh\ 
for halo stars. Some studies (Ryde \& Lambert \cite{ryde04},
Nissen et al. \cite{nissen04})
indicate that \sfe\ is approximately constant at a level of +0.3~dex in
the metallicity range $-3 < \feh < -1$. 
Such a plateau-like overabundance of S with respect to Fe is also
predicted from Galactic chemical evolution models with Type II SNe
as the dominant source of element production up to $\feh \sim -1$
(Chiappini et al. \cite{chiappini99},
Goswami \& Prantzos \cite{goswami00}); the decline of
\sfe\ for galactic disk stars (Chen et al. \cite{chen02}, Ryde \cite{ryde06})  
is then due to the release of iron from Type Ia SNe at $\feh > -1$. 

Other investigations point, however, to an increasing trend of \sfe\
towards lower metallicities (Israelian \& Rebolo \cite{israelian01};
Takada-Hidai et al. \cite{takada02}) with \sfe\ reportedly reaching as high as
+0.8~dex at $\feh \sim -2$. As discussed by Israelian \& Rebolo,
such high values of \sfe\ may be explained if 
SNe with very large explosion energies of $E = (10-100) \times 10^{51}$\,ergs
(so-called hypernovae) make a substantial contribution to the
nucleosynthesis of elements in the early Galaxy (Nakamura et al.
\cite{nakamura01}). An alternative explanation of an
increasing trend of \sfe\ towards lower metallicities has been 
proposed by Ramaty et al. (\cite{ramaty00}): assuming a short
mixing time ($\sim 1$~Myr) for supernovae-synthesized volatile elements
like oxygen and sulphur and a longer mixing time ($\sim 30$~Myr)
for refractory elements like Fe, high values
of \ofe\ and \sfe\ are expected in the early Galaxy. 

In a recent paper by Caffau et al. (\cite{caffau05}), more puzzling
data on \sfe\ in halo stars have been obtained. Both high 
$\sfe \sim +0.8$~dex and low $\sfe \sim +0.3$~dex are found in the
metallicity range $-2.2 < \feh < -1.0$ (see Fig. 10 of Caffau et al.),
suggesting a dichotomy of \sfe . If real, this
points to a very complicated chemical evolution of sulphur
in the early Galaxy.  

Additional problems have been revealed by Takeda et al.
(\cite{takeda05}), who find that S abundances derived from the weak
$\lambda 8694.6$ \SI\ line are systematically higher than S 
abundances derived from the stronger $\lambda \lambda 9212.9,9237.5$ pair 
of lines when their new non-LTE corrections are applied.

The uncertainty about the trend of \sfe\ calls for further
studies of sulphur abundances in halo stars. In this paper we 
present new sulphur abundances for 12 halo stars with $-3.3 < \feh < -1.9$
based on high-quality, near-IR spectra obtained with the VLT/UVES spectrograph.
We have also re-analyzed UVES spectra of 28 stars with $\feh < -1$
from Nissen et al. (\cite{nissen04}, hereafter Paper\,I) determining
\teff\ from the H$\beta$ line in the same way as for the new stars
and improving the derivation of S abundances from the
$\lambda \lambda 9212.9,9237.5$ lines by taking into account the opacity
contribution from the wings of the Paschen-zeta hydrogen line
at 9229\,\AA . In addition, possible non-LTE effects are investigated
by comparing S abundances obtained from the weak \SI\ line at 
8694.6\,\AA\ with data from the $\lambda \lambda 9212.9, 9237.5$ \SI\ lines.

We have also derived zinc abundances
from the $\lambda \lambda 4722.2,4810.5$ \ZnI\ lines.
Both S and Zn are among the few elements which are not
readily depleted onto dust in the interstellar medium of the
Milky Way and are present in the gas-phase in near-solar
proportions. For this reason, they are key to studies of
metal enrichment in distant galaxies, particularly those
detected as damped Lyman-alpha systems (DLAs) in the
spectra of high redshift QSOs.
Assuming that sulphur behaves like other
$\alpha$-capture elements and that Zn follows Fe, the S/Zn ratio may
be used to date the star formation process in DLAs 
(Wolfe et al. \cite{wolfe05}). A clarification
of the trends of both S and Zn for Galactic stars is important to test
if these assumptions are correct.

\section{Observations and data reduction}
\label{sect:obs}
The 12 new programme stars were selected from the list of very metal-poor
turnoff stars in Ryan et al. (\cite{ryan99}) and observed with the VLT/UVES
spectrograph (Dekker et al. \cite{dekker00}) in service mode during ESO
period 73 (April -September, 2004).
In addition to the programme stars, a fast-rotating bright early-type
B star (either \object{HR\,5488} or \object{HR\,6788}) was observed on 
each night in order to
be able to remove telluric \water\ lines in the 9212 - 9238~\,\AA\ region.

The UVES setup was the same as described in Paper\,I
except that no image slicer was applied when imaging 
the star onto the 0.7 arcsec wide entrance slit. Briefly, we mention
that the dichroic mode of UVES was used to cover the spectral region
$3750 - 5000$\,\AA\ in the blue arm and $6700 - 10500$\,\AA\ in the
red arm, in both cases with a spectral resolving power of
$\lambda/\Delta\lambda \simeq 60\,000$.
The blue region contains
a number of \FeII\ lines suitable for determining the iron abundance,
and two \ZnI\ lines.
In the red region we find the weak $\lambda$8694.6 \SI\ line 
as well as the stronger \SI\ lines at 9212.9 and 9237.5~\,\AA .
Typical S/N ratios are 300
in the blue spectral region, 250--300 at the $\lambda 8694.6$ \SI\ line,
and 150--200 in the 9212 -- 9238 \AA\ region.

\begin{figure}
\resizebox{\hsize}{!}{\includegraphics{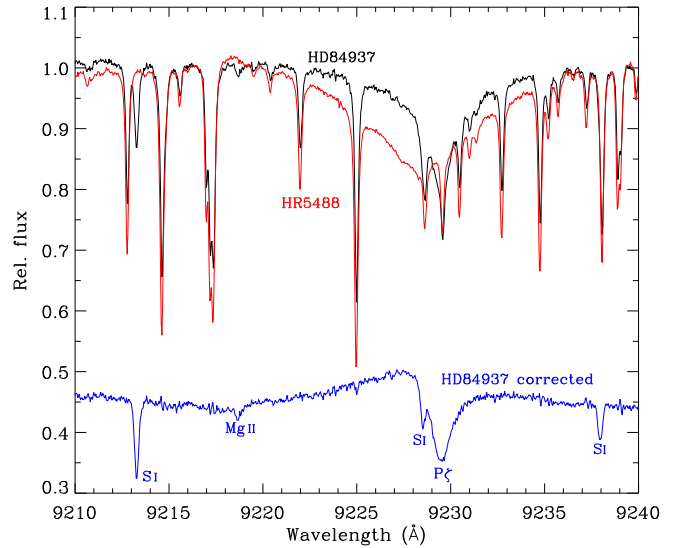}}
\caption{The VLT/UVES spectrum of \object{HD\,84937} in the spectral
region 9210 -- 9240\,\AA\
overlaid with the B2III spectrum of \object{HR\,5488}. Below
is the spectrum of \object{HD\,84937}
(shifted $\sim 0.5$ units in relative flux)
after removal of the telluric lines using the IRAF
task {\tt telluric} to divide with a scaled version of the spectrum
of \object{HR\,5488}.}.
\label{fig:telluric}
\end{figure}

The spectra were reduced by using standard IRAF routines for
order definition, background subtraction, flat-field correction,
order extraction with sky subtraction and wavelength calibration.
Continuum fitting was performed with the IRAF task {\tt continuum} using
spline functions with a scale of a few \AA . Since many continuum
windows are available for these metal-poor stars, even in the blue
part of the spectrum, this method works well.  

The quality of the 2004 spectra is about the same as shown in Fig. 2 of 
Paper\,I. As an additional illustration of
the removal of telluric lines, we show the spectrum of \object{HD\,84937} in
Fig.~\ref{fig:telluric}. This bright star was observed with particularly
high S/N. As seen from the figure, the  $\lambda 9237.5$ \SI\ line is
completely overlapped by a telluric line, but after division
with the scaled B-type spectrum of \object{HR\,5488}, the line emerges
very clearly. Due to the radial velocity difference of the two stars,
the \pzeta\ \HI\ line at 9229\AA\ is seen as a broad feature
in `emission' from \object{HR\,5488} overlaid with a narrower 
absorption line from \object{HD\,84937}. The wings
of the \pzeta\ line are lost and therefore the equivalent widths of the \SI\
lines can only be measured relative to the local continuum. Hence,
when using their equivalent widths to derive S abundances, the opacity
contribution from the \pzeta\ line should be taken into account.

\begin{figure}
\resizebox{\hsize}{!}{\includegraphics{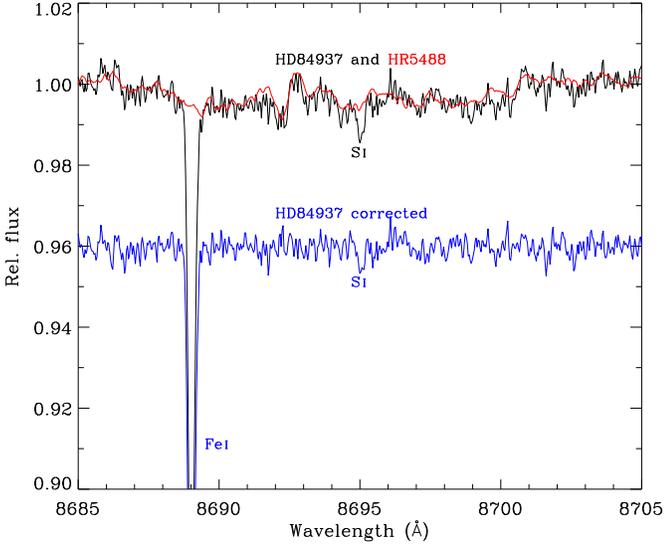}}
\caption{The VLT/UVES spectrum of \object{HD\,84937} in the spectral
region 8685 -- 8705\,\AA\
overlaid with the B2III spectrum of \object{HR\,5488}. Below
is the spectrum of \object{HD\,84937} (shifted down 0.04 units  in relative flux)
after division with the spectrum of \object{HR\,5488}.}
\label{fig:8694}
\end{figure}

Interference fringing is another problem in 
the near-IR part of the spectrum. The fringes on the UVES MIT CCD
have an amplitude of 20 - 40\,\%  at 9000\,\AA . After
flat-fielding, residual
fringes at a level of $\pm 0.5$\,\% remain as seen from Fig.~\ref{fig:8694}.
As discussed by Korn \& Ryde (\cite{korn05}), this makes it
difficult to determine precise S abundances of halo
stars from the weak $\lambda 8694.6$ S\,{\sc i} line.
B-type stars may, however, also be used to correct for the residual fringing
as illustrated in Fig.~\ref{fig:8694}. After division by the spectrum
of \object{HR\,5488}, the spectrum of \object{HD\,84937} has a S/N $\sim
500$. Without such a correction, the error of the S abundance derived
from the $\lambda 8694.6$ line can be large, especially for stars
with $\feh < -2$.

Equivalent widths of S, Fe and Zn lines
were measured by Gaussian fitting or direct integration
if the fit was poor and are given in Table \ref{tab:A.1} for
the 12 stars observed in 2004. The corresponding table for
the Paper\,I stars is available at the 
CDS\footnote{\tt http://cdsweb.u-strasbg.fr/cgi-bin/qcat?/A+A/415/993}.

\section{Model atmospheres and stellar parameters}
\label{parameters}
The determination of the effective temperature (\teff ) as well as 
the abundances of S, Fe and Zn is based on $\alpha$-element enhanced
(\alphafe = +0.4, $\alpha$ = O, Ne, Mg, Si, S, Ca, and Ti) 
1D model atmospheres computed with the Uppsala MARCS code.
Updated continuous opacities (Asplund et al. \cite{asplund97})
including UV line blanketing are used. 
LTE is assumed both in constructing the models and in deriving
\teff\ and the abundances. Convection is treated in the
approximation of Henyey et al. (\cite{henyey65})
with a mixing-length parameter of $\alpha_{\rm MLT} = l/H_p = 1.5$, and 
a temperature distribution in the convective elements 
determined by the diffusion equation (parameter
$y = 3/4\,\pi^2 = 0.076$). 
The reader is referred to Ludwig et al. (\cite{ludwig99}) for a summary of 
the parameters used in different versions of the mixing length theory. 

\subsection{Effective temperature}
\label{teff}
Instead of deriving \teff\ from
the \by\ and \VK\ colour indices as in Paper\,I, we decided to 
use the profile of the H$\beta$ line
in the blue part of the UVES spectra. The advantage is that
the H$\beta$-based \teff\ is not affected by possible errors
in the interstellar reddening; in general the new stars are more
distant than the stars in Paper\,I and hence potentially more affected
by interstellar absorption.

The H$\beta$ line is well centered in an echelle order having a
width of approximately 80\,\AA . After flat-fielding, the continua
of the two adjacent echelle orders were fitted by low order
spline functions and the continuum of the H$\beta$ order was determined
by interpolation between these functions in pixel space. The H$\beta$
echelle order is not wide enough to reach the true continuum, but the
method allows one to use the inner $\pm 30$\,\AA\ of the profile to
derive \teff\ from a comparison with synthetic H$\beta$ profiles for
a grid of model atmospheres. In addition to the 12 new stars, 
\teff\ was also re-determined in this way for stars from Paper\,I.

\begin{figure}
\resizebox{\hsize}{!}{\includegraphics{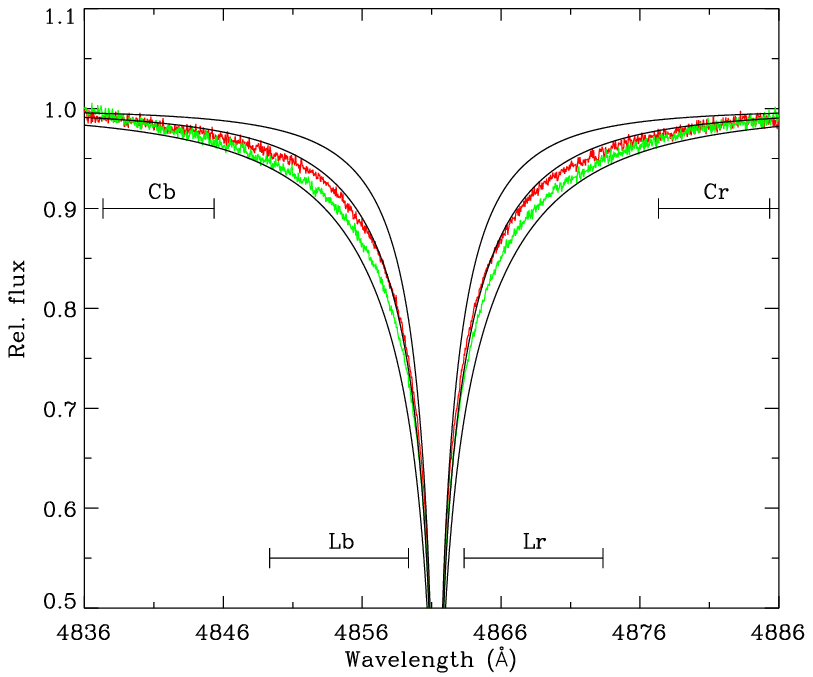}}
\caption{Observed H$\beta$ profiles in the spectra of
\object{CD\,$-24\degr17504$} (\teff = 6340\,K, upper jagged line)
and \object{LP\,815-43} (\teff = 6480\,K, lower jagged line) 
compared to synthetic
profiles (thin lines) computed for model atmospheres with
\logg\ = 4.3, $\feh = -2.8$ and \teff\ = 6000, 6300, 6600\,K, respectively.
An index, $\beta$(UVES), measuring the strength of the H$\beta$ line is defined
from the total flux in the C bands relative to the total flux in the L bands
(see Appendix \ref{appendix:B}).}
\label{fig:H-beta}
\end{figure}

To illustrate this method,
Fig.~\ref{fig:H-beta} shows the H$\beta$ profiles in the spectra of two stars
compared to synthetic profiles calculated as described
in Barklem et al. (\cite{barklem02}) using Stark broadening
from Stehl\'{e} \& Hutcheon (\cite{stehle99}) and self-broadening from 
Barklem et al. (\cite{barklem00a}).
A few metallic absorption lines have been removed by interpolating
the H$\beta$ profile across these lines. An index, $\beta$(UVES),
defined as the ratio of the total
flux in the two C-bands to the total flux in the two L bands 
is used to determine \teff\ (see Appendix \ref{appendix:B}).
Hence, we are not using the position of the
true continuum in determining \teff . The center of the H$\beta$ line is also 
avoided, because the fit between theoretical and observed profiles is
poor around the line center due to non-LTE effects 
(Przybilla \& Butler \cite{przybilla04}). 

For stars with $\feh > -1$, the metal lines in
the H$\beta$ line are so strong that the removal of 
these lines by interpolation of the profile across the lines
is unreliable. Hence, six stars with $\feh > -1$ from
Paper\,I are not included in the present paper.

The symmetric appearance of the observed profiles shows that the
procedure of rectifying the H$\beta$ echelle order has worked well.
In fact, the difference between \teff\ determined from the left
and the right part of the H$\beta$ profile never exceeded 30\,K.
The internal stability of the method is also very good; the effective
temperatures determined for a given star observed on different nights
agreed within $\pm 20$\,K. Another advantage is that the profile
of the H$\beta$ line varies very little with gravity and metallicity.
Hence, 
the errors in these parameters
have only a small effect on the derived \teff . For a group of stars
with similar atmospheric parameters, like metal-poor turnoff stars,
one might therefore expect that differential values of \teff\ can be 
determined with a precision of about  $\pm 30$\,K from the H$\beta$ profile. 

The systematic error of \teff\ is, however, larger.
The fit between the theoretical and observed profiles is
not perfect in the central region of the H$\beta$ line, and the estimated
value of \teff\ therefore depends to some extent on where the L-bands 
of the H$\beta$ index ½are placed.
The temperature structure of the model atmospheres also plays a role. 
If the mixing-length parameter is decreased from $l/H_p = 1.5$
to 1.0, the derived \teff\ changes by about $-25$\,K for metal-poor
stars at the turnoff ($\teff \sim 6400$\,K) and about $-60$\,K
for cooler stars ($\teff \sim 5800$\,K).

\begin{figure}
\resizebox{\hsize}{!}{\includegraphics{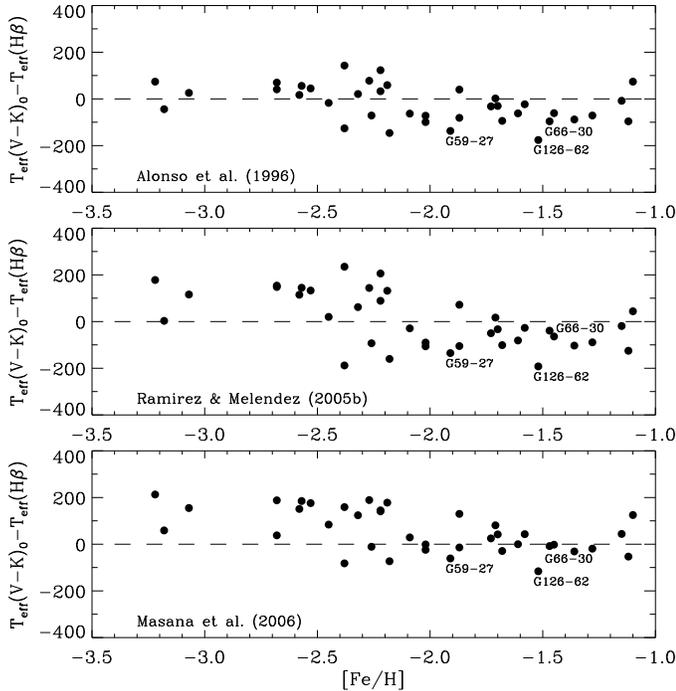}}
\caption{The difference of effective temperatures derived from \VKO\
and H$\beta$ using three different calibrations of \VKO .
The comparison includes only 38 stars; \object{LP\,651-4}
falls outside the range in \VKO\  where the calibrations are
valid, and \object{CD\,$-71\degr1234$} is missing Str\"{o}mgren photometry;
hence, its interstellar reddening excess, $E\VK$, could not be derived.
Three stars known to be single-lined spectroscopic binaries are marked.}
\label{fig:Teff_comp}
\end{figure}

The derived effective temperatures are given in Table \ref{table:par}.
Ten of the stars are
in common with Asplund et al. (\cite{asplund06}), who determined \teff\
from the H$\alpha$ profile in independent UVES spectra. The average difference,
\teff\,(H$\beta$) $-$ \teff\,(H$\alpha$) is 64\,K with a rms scatter of the
deviation of $\pm 28$\,K. This small scatter confirms that precise 
differential values of \teff\ can be determined from H$\alpha$ and H$\beta$ 
for stars with similar atmospheric parameters,
but the systematic difference indicates that the absolute values of \teff\
are more uncertain. Interestingly, the systematic difference of
\teff\,(H$\beta$) and \teff\,(H$\alpha$) vanishes if the 
mixing-length parameter is decreased to $l/H_p = 0.5$. The same 
conclusion was reached by Fuhrmann et al. (\cite{fuhrmann93}) 
for a set of Kurucz model atmospheres with the convective structure
parameter $y = 0.5$. Given that Barklem et al. (\cite{barklem02})
were not able to obtain a satisfactory fit to the H$\alpha$ and H$\beta$
line profiles in the solar spectrum for any combination of the
convection parameters $l/H_p$ and $y$, it seems, however, premature to change
these parameters from the standard values of the MARCS models. 
Clearly, the model atmospheres and/or the line broadening theory need to be improved.
Although it is often argued in the literature that the wings of Balmer lines
are formed in LTE for late-type stars, very recent work suggests that this
may in fact not necessarily be true (Barklem \cite{barklem07}). 
Unfortunately, the degree of departures from LTE depends critically
on still poorly known inelastic collisions with other neutral H atoms.
Some existing atomic calculations imply LTE while others yield
significant non-LTE effects.
A full 3D non-LTE analysis of Balmer line profiles in cool stars
with improved collisional data
is urgently needed to resolve this potentially significant systematic error.

The problem of systematic errors in the effective temperatures
of metal-poor stars is also evident when comparing \teff\,(\hbeta )
with effective temperatures derived from the \VKO\ index using
three different colour-\teff\ calibrations (see Fig.\,\ref{fig:Teff_comp}).
Here, the $V$ magnitude was taken from Str\"{o}mgren photometry
(see Table \ref{tab:B.1}), and $K$ from the 2MASS catalogue (Skrutski et al.
\cite{skrutskie06}). Correction for reddening was applied according
to the relation $E\VK = 3.8\,\,E\by$ (Savage \& Mathis \cite{savage79}),
where $E\by$ is derived as described in Appendix \ref{appendix:B}. 

The \VKO\ calibrations of Ram\'{\i}rez \& Melend\'{e}z (\cite{ramirez05b})
and Masana et al. (\cite{masana06}) refer directly to the 2MASS 
$K_s$ magnitude system, whereas the Alonso et al. (\cite{alonso96})
calibration refer to the original Johnson (\cite{johnson66})
$K_J$ system. Hence, before applying the  Alonso et al. calibration,
we converted $K_s$ to $K_J$ using 
equations in Alonso et al. (\cite{alonso94}) relating $K_J$ to
$K_{CIT}$ (Elias et al. \cite{elias82}) and the transformation
from $K_s$ to $K_{CIT}$ derived by Cutri et al. (\cite{cutri06}).
These transformations all contain a small \JK\ colour term, but
since our sample of stars are confined to a narrow
range in $J-K$, we get in a good approximation $(V\!-\!K_J) \simeq (V\!-\!K_s)
- 0.035$ for metal-poor stars in the turnoff region. If this transformation 
had not been applied, the effective temperatures derived from \VKO\
with the Alonso et al. calibration would have been about 65\,K lower.

\begin{table*}
\caption[ ]{Derived stellar parameters and abundances of sulphur and zinc.
LTE sulphur abundances are given for each of the three \SI\ lines. The LTE Zn abundance is
the average of the values derived from the $\lambda\lambda 4722.2, 4810.5$ \ZnI\ lines.
In calculating \sfe\ and \znfe , solar abundances \logFesun\ = 7.50, \logSsun\ = 7.20
and \logZnsun\ = 4.60 have been adopted. The non-LTE values of \sfe\ and \znfe\ are
based on the non-LTE corrections calculated by Takeda et al. (\cite{takeda05}) for
a hydrogen collisional parameter $S_{\rm H}$ = 1.}
\label{table:par}
\setlength{\tabcolsep}{0.13cm}
\begin{tabular}{lcccccclcccrr}
\noalign{\smallskip}
\hline\hline
\noalign{\smallskip}
  ID   &   \teff & \logg &  \feh & \micro & \multicolumn{3}{c}{\logSlte} & \sfe & \sfe & \logZn & \znfe & \znfe \\
       &   (K)   & (cgs) &       & (\kmprs) & $\lambda 9212.9$ & $\lambda 9237.5 $ & $\lambda 8694.6$ & LTE  & 
 non-LTE & LTE    & LTE      & non-LTE  \\
\noalign{\smallskip}
\hline
\noalign{\smallskip}
Paper\,\,I stars    &         &     &  & & & & & & & & &      \\
\noalign{\smallskip}
\object{BD\,$-13\degr3442$} & 6366 &   3.99 &  $-$2.69 &   1.5 &   4.92 &   4.95 &        &   0.43 &   0.23 &   1.99 &   0.08 &   0.19  \\
\object{CD\,$-30\degr18140$}& 6272 &   4.12 &  $-$1.89 &   1.5 &   5.59 &   5.68 &   5.52 &   0.29 &   0.21 &   2.74 &   0.03 &   0.10  \\
\object{CD\,$-35\degr14849$}& 6294 &   4.26 &  $-$2.34 &   1.5 &   5.14 &   5.16 &        &   0.29 &   0.18 &   2.33 &   0.07 &   0.14  \\
\object{CD\,$-42\degr14278$}& 6085 &   4.39 &  $-$2.03 &   1.5 &   5.35 &   5.40 &        &   0.20 &   0.14 &   2.65 &   0.08 &   0.12  \\
\object{G\,11-44}        &    6178 &   4.35 &  $-$2.03 &   1.5 &   5.54 &   5.49 &        &   0.34 &   0.27 &   2.76 &   0.19 &   0.24  \\
\object{G\,13-09}        &    6343 &   4.01 &  $-$2.29 &   1.5 &   5.38 &        &        &   0.47 &   0.31 &   2.40 &   0.09 &   0.18  \\
\object{G\,18-39}        &    6093 &   4.19 &  $-$1.46 &   1.5 &   6.05 &   6.09 &   6.08 &   0.33 &   0.26 &   3.22 &   0.08 &   0.13  \\
\object{G\,20-08}        &    6194 &   4.29 &  $-$2.19 &   1.5 &   5.26 &   5.40 &        &   0.32 &   0.24 &   2.56 &   0.15 &   0.21  \\
\object{G\,24-03}        &    6084 &   4.23 &  $-$1.62 &   1.5 &   5.98 &   5.91 &   5.89 &   0.34 &   0.29 &   3.04 &   0.06 &   0.11  \\
\object{G\,29-23}        &    6194 &   4.04 &  $-$1.69 &   1.5 &   5.76 &   5.79 &   5.84 &   0.29 &   0.20 &   2.86 &  $-$0.05 &   0.02  \\
\object{G\,53-41}        &    5993 &   4.22 &  $-$1.29 &   1.3 &   6.18 &   6.16 &   6.19 &   0.27 &   0.20 &   3.33 &   0.02 &   0.06  \\
\object{G\,64-12}        &    6435 &   4.26 &  $-$3.24 &   1.5 &   4.41 &        &        &   0.45 &   0.19 &   1.77 &   0.41 &   0.52  \\
\object{G\,64-37}        &    6432 &   4.24 &  $-$3.08 &   1.5 &   4.45 &        &        &   0.33 &   0.10 &   1.96 &   0.44 &   0.55  \\
\object{G\,66-30}$^{\rm a)}$& 6470 &   4.29 &  $-$1.48 &   1.5 &   6.03 &   6.11 &   5.84 &   0.28 &   0.18 &   3.10 &  $-$0.02 &   0.05  \\
\object{G\,126-62}$^{\rm a)}$&6224 &   4.11 &  $-$1.55 &   1.5 &   5.95 &   5.96 &   5.78 &   0.25 &   0.16 &   3.04 &  $-$0.01 &   0.05  \\
\object{G\,186-26}       &    6417 &   4.42 &  $-$2.54 &   1.5 &   4.85 &   4.85 &        &   0.19 &   0.07 &   2.25 &   0.19 &   0.26  \\
\object{HD\,106038}      &    6027 &   4.36 &  $-$1.37 &   1.2 &   6.28 &   6.22 &   6.29 &   0.43 &   0.38 &   3.37 &   0.14 &   0.18  \\
\object{HD\,108177}      &    6156 &   4.28 &  $-$1.71 &   1.5 &        &   5.86 &   5.78 &   0.33 &   0.30 &   2.96 &   0.07 &   0.12  \\
\object{HD\,110621}      &    6157 &   4.08 &  $-$1.59 &   1.5 &   6.03 &   5.92 &   5.74 &   0.29 &   0.21 &   3.02 &   0.01 &   0.07  \\
\object{HD\,140283}      &    5849 &   3.72 &  $-$2.38 &   1.5 &   5.07 &   5.06 &   5.17 &   0.28 &   0.20 &   2.30 &   0.08 &   0.14  \\
\object{HD\,160617}      &    6047 &   3.84 &  $-$1.75 &   1.5 &   5.81 &   5.80 &   5.85 &   0.37 &   0.28 &   2.82 &  $-$0.03 &   0.04  \\
\object{HD\,179626}      &    5881 &   4.02 &  $-$1.12 &   1.4 &   6.37 &        &   6.31$^{\rm b)}$ &   0.26 &   0.17 &   3.48 &   0.00 &   0.04  \\
\object{HD\,181743}      &    6044 &   4.39 &  $-$1.87 &   1.5 &   5.62 &   5.62 &        &   0.29 &   0.23 &   2.81 &   0.08 &   0.12  \\
\object{HD\,188031}      &    6234 &   4.16 &  $-$1.72 &   1.5 &   5.76 &   5.75 &   5.78 &   0.28 &   0.21 &   2.91 &   0.03 &   0.09  \\
\object{HD\,193901}      &    5699 &   4.42 &  $-$1.10 &   1.2 &   6.29 &   6.30 &   6.34 &   0.21 &   0.17 &   3.38 &  $-$0.12 &  $-$0.10  \\
\object{HD\,194598}      &    6020 &   4.30 &  $-$1.15 &   1.4 &   6.26 &   6.30 &   6.20$^{\rm b)}$ &   0.20 &   0.13 &   3.37 &  $-$0.08 &  $-$0.04  \\
\object{HD\,215801}      &    6071 &   3.83 &  $-$2.28 &   1.5 &   5.28 &   5.16 &        &   0.30 &   0.17 &   2.41 &   0.09 &   0.17  \\
\object{LP\,815-43}      &    6483 &   4.21 &  $-$2.71 &   1.5 &   4.87 &   4.81 &        &   0.35 &   0.17 &   2.17 &   0.28 &   0.38  \\
\noalign{\smallskip}
New stars   &         &     &  & & & & & & & & &     \\
\noalign{\smallskip}
\object{CD\,$-24\degr17504$} &6338 &   4.32 &  $-$3.21 &   1.5 &   4.51 &        &        &   0.52 &   0.30 &   1.96 &   0.57 &   0.66  \\
\object{CD\,$-71\degr1234$}$^{\rm a)}$ & 6325 &   4.18 &  $-$2.38 &   1.5 &   5.13 &   5.19 &        &   0.34 &   0.22 &   2.34 &   0.12 &   0.20  \\
\object{CS\,22943-095}   &    6349 &   4.18 &  $-$2.24 &   1.5 &   5.36 &   5.33 &        &   0.38 &   0.26 &   2.49 &   0.13 &   0.20  \\
\object{G\,04-37}        &    6308 &   4.25 &  $-$2.45 &   1.5 &   5.05 &   4.98 &        &   0.26 &   0.15 &   2.29 &   0.14 &   0.21  \\
\object{G\,48-29}        &    6482 &   4.25 &  $-$2.60 &   1.5 &   4.94 &   5.03 &        &   0.39 &   0.23 &   2.13 &   0.13 &   0.23  \\
\object{G\,59-27}$^{\rm a)}$& 6272 &   4.23 &  $-$1.93 &   1.5 &   5.57 &   5.57 &        &   0.30 &   0.20 &   2.73 &   0.06 &   0.12  \\
\object{G\,126-52}       &    6396 &   4.20 &  $-$2.21 &   1.5 &   5.21 &        &        &   0.22 &   0.09 &   2.50 &   0.11 &   0.19  \\
\object{G\,166-54}       &    6407 &   4.28 &  $-$2.58 &   1.5 &   5.04 &   5.06 &        &   0.43 &   0.29 &   2.10 &   0.08 &   0.16  \\
\object{HD\,84937}       &    6357 &   4.07 &  $-$2.11 &   1.5 &   5.47 &   5.38 &   5.40 &   0.33 &   0.23 &   2.55 &   0.06 &   0.15  \\
\object{HD\,338529}      &    6373 &   4.03 &  $-$2.26 &   1.5 &   5.36 &   5.32 &        &   0.40 &   0.25 &   2.48 &   0.14 &   0.23  \\
\object{LP\,635-14}      &    6367 &   4.11 &  $-$2.39 &   1.5 &   5.12 &   5.07 &        &   0.29 &   0.14 &   2.36 &   0.15 &   0.24  \\
\object{LP\,651-4}       &    6371 &   4.20 &  $-$2.63 &   1.5 &   4.83 &        &        &   0.26 &   0.11 &   2.04 &   0.07 &   0.16  \\
\hline
\end{tabular}
\begin{list}{}{}
\item[$^{\rm a)}$] Single-lined spectroscopic binary star.
\item[$^{\rm b)}$] The weaker $\lambda 8694.0$ \SI\ line is included in
the determination of this S abundance.
\end{list}
\end{table*}

As seen from Fig.\,\ref{fig:Teff_comp}, there is good agreement
between \teff\ from \hbeta\ and from \VKO\ when using the
Alonso et al. (\cite{alonso96}) calibration. Excluding the three known binary
stars, the mean difference,
$\teff\,\VKO\,-\,\teff\,(\hbeta)$ is $-11 \pm 72$\,K. The scatter
agrees well with the error of $\teff\,\VKO$ expected from errors
in the reddening correction. Still, there is a tendency for
the residuals to increase with decreasing \feh . This tendency is even more
pronounced for the \VKO\ calibrations of
Ram\'{\i}rez \& Melend\'{e}z (\cite{ramirez05b}) and
Masana et al. (\cite{masana06}).
When using the \VKO\ calibration of 
Ram\'{\i}rez \& Melend\'{e}z, stars with
$\feh > -2.0$ have a mean deviation $\teff\,\VKO\,-\,\teff\,(\hbeta)$
of about $-50$\,K, whereas stars with 
$\feh < -2.5$ have a mean deviation of +124\,K.
Hence, our H$\beta$-based effective temperatures do not confirm the
hot \teff\ scale of very metal-poor turnoff stars derived by
Ram\'{\i}rez \& Melend\'{e}z (\cite{ramirez05a}) from their
application of the infrared flux method. 

Four of the stars in Table \ref{table:par} are known to be
single-lined spectroscopic
binaries, i.e. \object{G\,66-30} (Carney et al. \cite{carney01}),
\object{G\,126-62} and \object{G\,59-27}  (Latham et al. \cite{latham02}),
and \object{CD$-71\degr1234$} (Ryan et al. \cite{ryan99}).
As discussed by Ram\'{\i}rez et al. (\cite{ramirez06}) in the case of
\object{G\,126-62}, these stars may have a cool companion that
causes too low a value of \teff\ to be derived from \VKO . Indeed, 
\object{G\,126-62} and \object{G\,59-27} exhibit some of the largest
deviations in Fig.\,\ref{fig:Teff_comp}, whereas \object{G\,66-30} does not
show a significant deviation. This star has an unusually high effective 
temperature (\teff = 6470\,K) for a halo star with \feh = $-1.48$,
and is classified as a blue straggler by Carney et al. (\cite{carney01}). 

\subsection{Surface gravity}
\label{sect:logg}
As in Paper\,I,
the surface gravity is determined from the fundamental relation
\begin{eqnarray}
\log \frac{g}{g_{\odot}}  =  \log \frac{\cal{M}}{\cal{M}_{\odot}} +
4 \log \frac{\teff}{T_{\rm eff,\odot}} + 
0.4 (M_{\rm bol} - M_{{\rm bol},\odot}) 
\end{eqnarray}
where $\cal{M}$ is the mass of the star and $M_{\rm bol}$ the absolute bolometric
magnitude.

The Str\"{o}mgren indices $(b-y)$, $m_1$ and $c_1$
were used to derive absolute visual magnitudes \Mv\ 
(see Appendix \ref{appendix:B}).
If the Hipparcos parallax (ESA \cite{esa97}) of the star
is available with an error $\sigma (\pi) / \pi < 0.3$, then \Mv\ was also
determined directly and averaged with the photometric value. The
bolometric correction was adopted from Alonso et al. (\cite{alonso95})
and the stellar mass derived by interpolating in the \Mv -- \logteff\
diagram between the $\alpha$-element enhanced evolutionary tracks of
VandenBerg et al. (\cite{vandenberg00}). As discussed in Paper\,I, 
the error of \logg\ is estimated to be about $\pm 0.15$\,dex.

For one star, \object{CD\,$-71\degr1234$}, neither $uvby$-$\beta$ photometry
nor the Hipparcos parallax is available. In this case, we estimated \logg\
by requiring that the difference in the Fe abundance derived from
\FeI\ and \FeII\ lines should equal the average difference between the two sets
of Fe abundances for the other stars (see Sect. \ref{sect:iron}).

The derived values of \teff\ and \logg\ are mutually dependent and are
also affected by \feh . Hence, the procedure of determining
atmospheric parameters and Fe abundances was iterated until consistency
was achieved. The final values obtained are given in Table~\ref{table:par}.

\section{Stellar abundances}
\label{sect:abundances}
Using MARCS model atmospheres with the parameters listed in
Table~\ref{table:par},
the  Uppsala {\sc eqwidth} and {\sc bsyn} programs were used to compute
equivalent widths of observed lines as a function of the corresponding
element abundance.
By interpolation to the observed equivalent width, the abundance of the
element is then derived. A basic assumption is LTE, but the effect
of possible departures from LTE will be discussed.

\subsection{Iron}
\label{sect:iron}
The Fe abundances are based on the \FeII\ lines listed in 
Table 2 of Paper\,I. The $gf$ values of these lines were 
derived by an inverted abundance
analysis of four `standard' stars, for which iron abundances were adopted 
from Nissen et al. (\cite{nissen02}). Hence, our \feh\ scale is tied
to that of Nissen et al., who used a set of very weak \FeII\ lines
in the red spectral region to determine differential Fe abundances with 
respect to the Sun. The $gf$ values agree very well with those 
recently derived by 
Melend\'{e}z et al. (\cite{melendez06}) from \FeII\ radiative lifetime 
measurements of Schnabel et al. (\cite{schnabel04}) and 
branching ratios computed by R.L. Kurucz. 
The average difference  of log\,$gf$  (Melend\'{e}z et al. $-$ Paper\,I) for the
19 \FeII\ lines is +0.04\,dex with a scatter of $\pm 0.06$\,dex only.  

Collisional line broadening was included in accordance with Barklem 
\& Aspelund-Johansson (\cite{barklem05}). For the large majority
of stars, the \FeII\ lines are so weak that the derived metallicity is
insensitive to possible errors in the van der Waals damping constant 
and also practically independent of
the microturbulence parameter. Stars with $\feh < -1.5$ are 
assumed to have \micro = 1.5\,\kmprs . For more metal-rich stars
\micro\ was determined by requesting that the derived 
\feh\ value should be independent of equivalent width. 

For all new stars, the Fe abundance was also determined from seven
unblended \FeI\ lines selected to be of similar strengths 
as the \FeII\ lines. Wavelengths, $gf$-values (adopted from O'Brian
et al. \cite{obrian91}) and measured equivalent widths are given in
Table \ref{tab:A.1}. Collisional broadening data were adopted from 
Barklem et al. (\cite{barklem00b}).

\begin{figure}
\resizebox{\hsize}{!}{\includegraphics{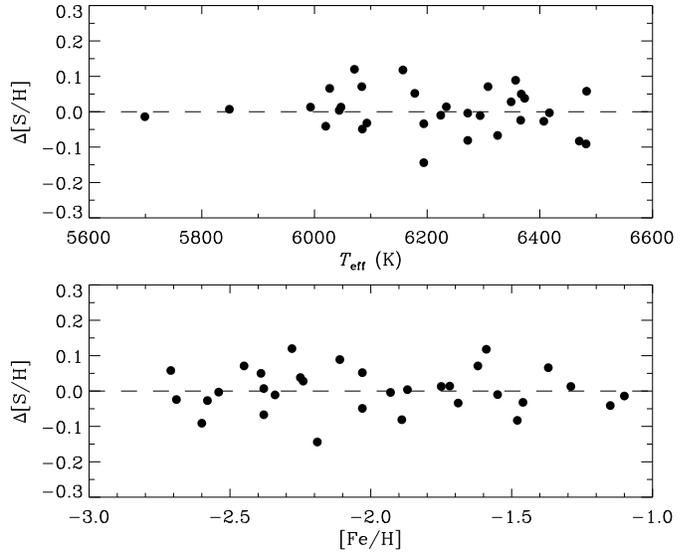}}
\caption{The difference in S abundances derived from
the $\lambda 9212.9$ and  $\lambda 9237.5$ \SI\ lines as a function of
\teff\ and \feh . The rms deviation
between the two sets of abundances is $\pm 0.06$\,dex.}
\label{fig:S.9212-9237}
\end{figure}

The average difference between Fe abundances 
derived from \FeII\ and \FeI\ lines is 0.14\,dex with a rms deviation
of $\pm 0.03$\,dex. This small scatter testifies to the high internal
precision obtained. The systematically higher Fe abundance obtained
from \FeII\ lines may be due to non-LTE effects on the \FeI\ lines
(see discussion in Asplund \cite{asplund05}),
although the scale of \teff\ and/or the  $gf$-values could also play a
role. We note that the \FeII\ - \FeI\ differences are very similar to those
estimated in Asplund et al. (\cite{asplund06}).
The \feh\ values given in Table \ref{table:par}
are based on the \FeII\ lines, and refer to an adopted
solar iron abundance of
\logFesun\ = 7.50, which is close to the value inferred from weak \FeII\
lines in the solar spectrum using a MARCS model for the Sun
(Nissen et al. \cite{nissen02}).

\subsection{Sulphur}
\label{sect:sulphur}
Sulphur abundances were derived from the observed equivalent widths of the
high excitation \SI\ lines at 8694.6, 9212.9 and 9237.5\,\AA . 
In two of the most metal-rich stars
(\object{HD\,179626} and \object{HD\,194598}), 
the $\lambda 8694.0$\,\SI\ line could also be detected, and hence
its corresponding S abundance was averaged with that from the 
$\lambda 8694.6$ line. The atomic data for the lines are given in Table 
\ref{tab:A.1}. The
$gf$-values (adopted from Coulomb approximation calculations of
Lambert \& Luck \cite{lambert78}) agree well with the experimental values of
Bridges \& Wiese (\cite{bridges67}). 
Somewhat different $gf$-values have been calculated by Bi\'{e}mont et al. 
(\cite{biemont93}) but, as noted in Paper\,I,
our adopted $gf$-values lead to a solar abundance \logSsun\ = 7.20, when
the $\lambda \lambda 8694.0,8694.6$ lines are analyzed with a MARCS model
of the Sun. This is close to the meteoritic abundance of sulphur,
\logS\ = $7.16 \pm 0.04$ (Asplund et al. \cite{asplund05}).

As mentioned in Sect \ref{sect:obs}, the equivalent widths of the \SI\
lines at 9212.9 and 9237.5\,\AA\ were measured relative to the flux in
the wings of the Paschen-zeta \HI\ line at 9229\AA . In Paper\,I, the
effect of \pzeta\ was neglected, but here we include
the opacity contribution from the wings when computing the 
equivalent widths. As in the case of \hbeta , a version
of the Uppsala {\sc bsyn} program kindly supplied by P. Barklem was used in 
these calculations. The maximum effect of \pzeta\ on the derived
S abundances occurs for the hottest turnoff stars and amounts to an
increase of \logS\ by about 0.07\,dex in the case of the 9237.5\,\AA\
line and about 0.02\,dex for the 9212.9\,\AA\ line. After inclusion of the
\pzeta\ opacity, there is an excellent agreement between
S abundances derived from the two lines with no significant trend of the 
deviation as a function of \teff\ or \feh\ (see Fig. \ref{fig:S.9212-9237}), in
contrast to the corresponding figure in Paper\,I, where a small but
significant trend with \teff\ was seen.

\begin{figure}
\resizebox{\hsize}{!}{\includegraphics{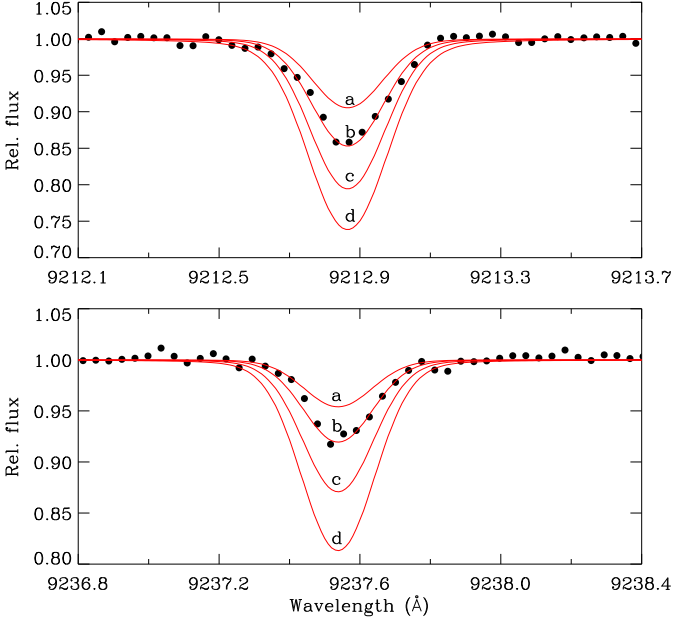}}
\caption{UVES observations (dots) of the $\lambda \lambda 9212.9, 9237.5$
\SI\ lines in the spectrum of HD\,181743.
The four synthetic profiles marked by $a$, $b$, $c$ and $d$ refer
to \logS = 5.33, 5.63, 5.93 and 6.23 corresponding to \sfe\ = 0.0, 0.3, 0.6
and 0.9, respectively.}
\label{fig:181743}
\end{figure}

The sulphur abundances derived from the equivalent widths of the three
\SI\ lines are given in Table \ref{table:par} together with
the average value of \sfe .
For some stars it was tested that a detailed
synthesis of the observed profiles of the \SI\ lines yields practically the
same abundances as the equivalent widths 
(see example in Fig. \ref{fig:181743}).

\begin{figure}
\resizebox{\hsize}{!}{\includegraphics{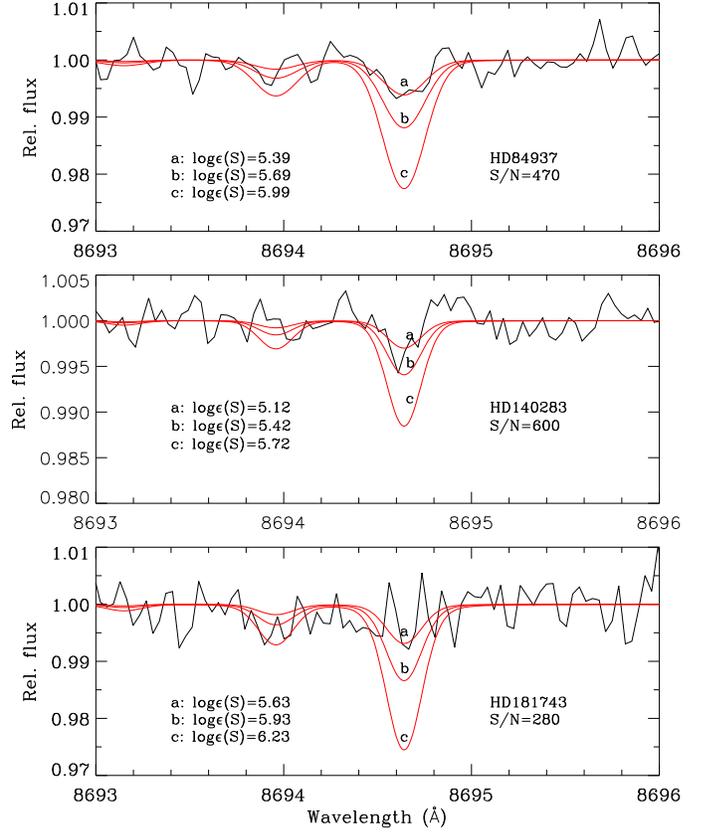}}
\caption{UVES spectra (jagged line) of the region around
the $\lambda 8694.6$ \SI\ line.
For each star, the three synthetic profiles marked by $a$, $b$ and
$c$ correspond to \sfe\ = 0.3, 0.6 and 0.9.}
\label{fig:syn8694}
\end{figure}

Table \ref{table:par} includes three bright halo stars, \object{HD\,84937}, 
\object{HD\,140283} and \object{HD\,181743}, claimed by others
to have very high sulphur abundances, $\sfe \simgt +0.6$, whereas we find
these stars to have \sfe\ around +0.3\,dex. In trying to
find the reason for this discrepancy, we briefly discuss these stars.
Fig. \ref{fig:syn8694} shows the spectrum synthesis of their 
\SI\ lines at 8694\,\AA . The
instrumental and stellar line broadening profile was 
approximated by a Gaussian with a FWHM of 8\,\kmprs\ in the case of 
\object{HD\,84937} and 6\,\kmprs\ for the other two stars. These
values were derived from fitting the \FeI\ line at 8688.6\,\AA .

\subsubsection{\object{HD\,84937}}
From a KECK/HIRES spectrum of \object{HD\,84937}, Takada-Hidai et al. 
(\cite{takada02}) measured the equivalent width of the $\lambda 8694.6$ 
\SI\ line to be $W_\lambda = 3.4$\,m\AA\ and derived \logS = 5.7 corresponding
to \sfe = +0.6. From our UVES spectrum we measure 
$W_\lambda =  1.6 \, \pm 0.3$\,m\AA , where the quoted (1-sigma) error
is estimated from the S/N of the spectrum and the uncertainty of the
continuum setting. The correspondingly derived sulphur abundance is
\logS\ = $5.40 \, \pm 0.08$.
As shown in Fig. \ref{fig:syn8694}, this agrees
well with the spectrum synthesis of the line; a value as high as 
\logS = 5.7 is clearly excluded. Furthermore, we obtain   
\logS = 5.39 and 5.48 from the $\lambda 9212.9, 9237.5$ \SI\  lines.
Altogether, our sulphur abundance of \object{HD\,84937} is about a 
factor of two lower than the value found by Takada-Hidai et al.
(\cite{takada02}).

\subsubsection{\object{HD\,140283}}
Using the UVES Paranal Observatory Project
(Bagnulo et al. \cite{bagnulo03}) spectrum of \object{HD\,140283},
Takeda et al. (\cite{takeda05}) derived $\sfe \sim +1.0$ (corresponding to
\logS $\sim 5.8$) from the $\lambda 8694.6$ \SI\ line (see Fig. 7
in Takeda et al. \cite{takeda05}). As seen from our Fig. \ref{fig:syn8694}, we 
barely detect this line in our S/N $\sim 600$ spectrum
of \object{HD\,140283}; clearly,
\logS\ is much lower than 5.8\,dex. From the measured equivalent
width of the $\lambda 8694.6$ line, 
$W_\lambda = 0.7 \, \pm 0.3$\,m\AA , we get \logS\ = $5.17 \, \pm 0.20$
in good agreement with the result from the 
$\lambda \lambda 9212.9, 9237.5$ \SI\ lines, i.e. \logS\ = 5.07 and
5.06, respectively. Judging from Fig. 7 in Takeda et al. (\cite{takeda05}),
the  UVES POP spectrum of \object{HD\,140283} is affected by residual
fringing, which in this case apparently has led to a spurious enhancement of the 
$\lambda 8694.6$ \SI\ line, as also pointed out by Korn \& Ryde (\cite{korn05}). 
Furthermore, it seems that Takeda et al.
have placed the continuum in the $\lambda 8694.6$ region too high.
Using the same UVES POP spectrum with a more reasonable continuum,
Takada-Hidai (\cite{takada05}) derived \logS\ = 5.53 from the 
$\lambda 8694.6$ line. This is, however,
still much higher than our value and also higher than the value of
\logS\ = 5.18, which they derive from the $\lambda 9237.5$ \SI\ line.

\subsubsection{\object{HD\,181743}}
For this star, Caffau et al. (\cite{caffau05}) obtained \logS = 6.23
and [S/Fe] = +0.84. The result is based on an
apparently clear detection of the $\lambda 8694.6$ \SI\ line in a
UVES spectrum with S/N = 180 (P. Bonifacio, private communication). However,
we cannot detect this line in our S/N $\sim 280$ spectrum as seen
from Fig. \ref{fig:syn8694}, but estimate a  3-sigma upper limit
of the equivalent width, $W_{\lambda} < 2$\,m\AA\ 
corresponding to \logS\ $< 5.8$. 
Caffau et al. did not use other \SI\ lines for this star, whereas 
we derive \logS = 5.62 from the $\lambda \lambda 9212.9, 9237.5$ lines. As seen 
from Fig. \ref{fig:181743} the fit of the corresponding synthetic 
profiles to the observed lines is excellent.

We note that the large differences between sulphur abundances
derived in the cited works and in the present paper cannot be explained
in terms of different values of \teff\ and \logg . In our view, it is 
more likely
that the S abundances derived by others from the weak $\lambda 8694.6$ \SI\ 
line have been overestimated due to fringing residuals in their spectra.
In general, more precise S abundances can be derived from the 
$\lambda \lambda 9212.9, 9237.5$ lines as they are more easily detected
and less susceptible to observational uncertainties. In this connection, it should
be noted that Caffau et al. find a few stars to have very high
\sfe\ values even on the basis of these lines. Most remarkable is 
\object{BD\,$+02\degr263$}, for which they get \sfe\ = +0.91. This star is,
however, a single-lined spectroscopic binary according to 
Latham et al. (\cite{latham02}), which may have affected the determination of 
\teff\ and \logg\ and hence the abundances of S and Fe.

\subsubsection{3D and non-LTE effects on the \SI\ lines}
\label{sect:non-LTE}
In Paper\,I it was shown that the application of 3D hydrodynamical model 
atmospheres instead of classical 1D models has a rather small effect 
on the derived sulphur abundance (see also Asplund \cite{asplund05}). 
For the large majority of our stars,
3D models lead to an
increase of \logS\ by 0.05 to 0.10\,dex. This
increase is, however, compensated by about the same increase in iron
abundance (see Table 5 in Paper\,I for details).
Hence, \sfe\ is practically unchanged.

\begin{figure}
\resizebox{\hsize}{!}{\includegraphics{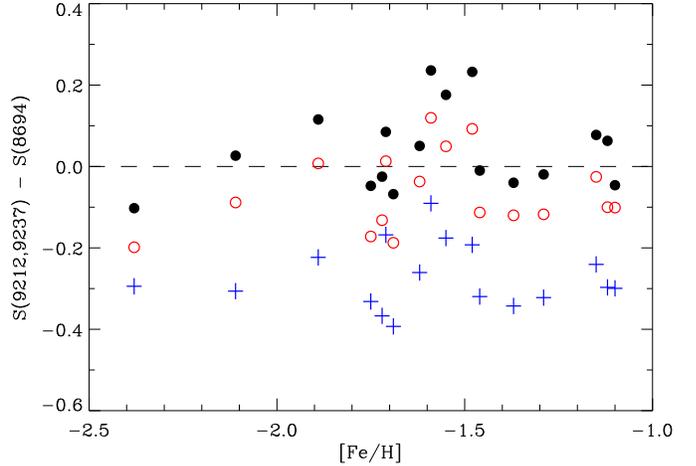}}
\caption{Comparison of S abundances derived from the \SI\
9212.9, 9237.5\,\AA\ pair and the $\lambda 8694.6$ line.
Filled circles refer to LTE. Open circles include non-LTE
corrections according to Takeda et al. (\cite{takeda05})
with the hydrogen collisional parameter $S_{\rm H} = 1$, and crosses refer
to $S_{\rm H} = 0.1$.}
\label{fig:non-LTE}
\end{figure}

It was noted in Paper\,I that the good agreement between S abundances
derived from the weak $\lambda 8694.6$ \SI\ line and the stronger
$\lambda \lambda 9212.9, 9237.5$ lines suggests that non-LTE effects on 
the sulphur lines are small. Thanks to the recent non-LTE calculations 
of Takeda et al.  (\cite{takeda05}), this statement may now be quantified. 
From extensive statistical equilibrium calculations Takeda et al. show that
the $\lambda \lambda 9212.9, 9237.5$ lines can suffer from significant
negative non-LTE corrections, whereas the non-LTE effects on the 
$\lambda 8694.6$ line are small. The size of the corrections depend
on the rate of inelastic collisions with neutral hydrogen, for which
Takeda et al. adopt the classical approximation of Drawin 
(\cite{drawin68}, \cite{drawin69}) in the version
of  Steenbock \& Holweger (\cite{steenbock84}) with a scaling
factor $S_{\rm H}$ that is varied from $10^{-3}$ to 10.

As seen from the tables of Takeda et al. (\cite{takeda05}), the non-LTE
effects vary quite strongly with \teff , \logg\ and \feh , but interpolation 
in the tables allows us to derive non-LTE corrections for our set of stars.
In Fig. \ref{fig:non-LTE}, the resulting difference of the S abundance
derived from the $\lambda 9212.9, 9237.5$\,\AA\ pair and the 
$\lambda 8694.6$ line has been plotted for two values of $S_{\rm H}$
and also in the LTE case. As seen, both LTE and $S_{\rm H} = 1$ give a 
satisfactory agreement between the two sets of S abundances, whereas the
case of $S_{\rm H} = 0.1$ can be excluded. The average difference and
dispersion of \logS $_{9212,9237}$ $-$ \logS $_{8694}$ are as follows: 
$+0.04 \pm 0.10$ (LTE), $-0.06 \pm 0.09$ (non-LTE with $S_{\rm H} = 1$),
and $-0.27 \pm 0.08$ (non-LTE with $S_{\rm H} = 0.1$). This suggests
that hydrogen collisions are quite efficient in thermalizing the 
\SI\ atoms. Still, the non-LTE effects for the
$\lambda \lambda 9212.9, 9237.5$ lines may be of importance for the
derived \sfe\ trend, due to varying non-LTE abundance corrections
as a function of \feh .
In the case of $S_{\rm H} = 1$, the corrections obtained by Takeda et al. 
range from about $-0.06$\,dex for the coolest
of our stars to about $-0.25$\,dex for the hottest and
most metal-poor stars (e.g. \object{G\,64-12}).

\subsubsection{CRIRES observations of the 1.046\,$\mu$m \SI\ triplet}
\label{CRIRES}
As part of the science verification of the ESO/VLT Cryogenic high-resolution
IR Echelle Spectrograph, CRIRES (K\"{a}ufl et al. \cite{kaufl04}),
a spectrum around the 1.046\,$\mu$m \SI\ triplet was obtained for
\object{G\,29-23} ($V = 10.19$, \feh $=-1.69$) on October 6, 2006.
The entrance slit width of
CRIRES was set at 0.4 arcsec, which corresponds to a resolution of 
$\lambda/\Delta\lambda  \simeq 50\,000$ with four detector pixels 
per spectral resolution bin $\Delta\lambda$.
In order to improve removal of sky emission and detector dark
current, the observations were performed in nodding mode with
a shift of 10 arcsec between the two settings of the star in
the slit. The exposure time
was 2400 sec. The seeing was rather poor (about 1.3 arcsec)
but adaptive optics was applied to improve the stellar image, and 
the combined spectrum has a very satisfactory S/N of 330
per spectral dispersion pixel. For comparison, we note that our UVES 
spectrum of \object{G\,29-23} (exposure time 1800 sec) has S/N $\sim 200$ 
around the \SI\ lines at 9212--9238\,\AA .\footnote{The UVES spectra
also cover the 1.046\,$\mu$m \SI\ triplet but due to the low efficiency
of the MIT CCD at this wavelength, the quality is far too low ($ S/N < 25$)
for a precise determination of the sulphur abundance.} Furthermore, 
unlike the UVES near-IR spectrum, the CRIRES spectrum at 1.046\,$\mu$m
is not plagued by telluric lines and fringing residuals.

\begin{table}
\caption[ ]{Atomic data for the 1.046\,$\mu$m \SI\ triplet.
Column 5 lists the equivalent width in the spectrum of
\object{G\,29-23} and the last column gives the derived
LTE sulphur abundance.}
\label{table:1046}
\setlength{\tabcolsep}{0.15cm}
\begin{tabular}{cccrcc}
\noalign{\smallskip}
\hline\hline
\noalign{\smallskip}
    & Wavelength & Exc.\,Pot. & log\,$gf$ & $W_\lambda$(G29-23) & \logS  \\
    & \AA        &  eV        &           &       m\AA        &        \\
\hline
\noalign{\smallskip}
\SI  & 10455.5  & 6.86 & 0.26    & 39.5 & 5.82 \\
\SI  & 10456.8  & 6.86 & $-$0.44 & 10.1 & 5.75 \\
\SI  & 10459.4  & 6.86 & 0.03    & 25.7 & 5.77 \\
\noalign{\smallskip}
\hline
\end{tabular}
\end{table}

\begin{figure}
\resizebox{\hsize}{!}{\includegraphics{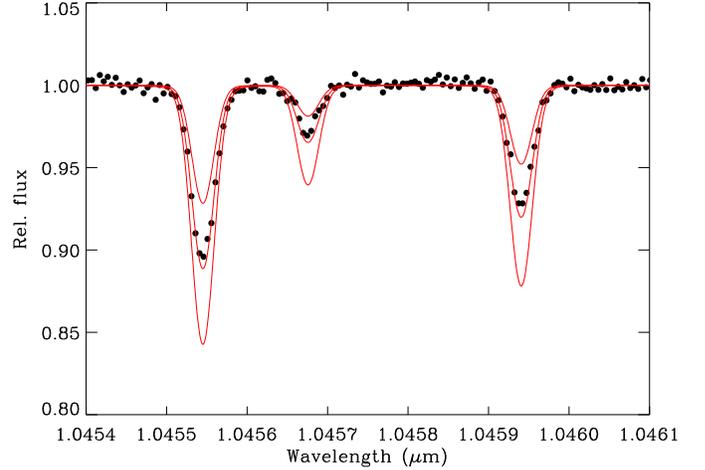}}
\caption{The CRIRES spectrum of \object{G\,29-23} around the
1.046\,$\mu$m \SI\ triplet (dots) compared with synthetic
profiles for three sulphur abundances, \logS\ = 5.51, 5.81 and 6.11
corresponding to \sfe\ = 0.0, 0.3 and 0.6, respectively.
As seen, the \sfe\ = 0.3 case provides an excellent fit to the observations.}
\label{fig:1046}
\end{figure}

The CRIRES data was reduced by using standard IRAF tasks  
for subtraction of bias and dark, spectrum extraction, and wavelength
calibration. Note that CRIRES supports high accuracy wavelength calibration
by means of a Th-Ar hollow cathode lamp (Kerber et al. \cite{kerber06}). 
Flatfielding was performed by the aid of the extracted 
spectrum of the bright ($V = 4.28$) B9III star \object{HD\,15315}. 
The resulting
spectrum of \object{G\,29-23} is shown in Fig. \ref{fig:1046}
and compared with a spectral synthesis of the three sulphur lines.

Atomic data for the 1.046\,$\mu$m \SI\ triplet are given in Table 
\ref{table:1046} together with the measured equivalent widths in the
\object{G\,29-23} spectrum. As in the case of the other \SI\ lines the
$gf$ values were taken from Lambert \& Luck (\cite{lambert78})
and collisional broadening data are from Barklem et al. (\cite{barklem00b}). 
The derived LTE S abundances given in Table \ref{table:1046} agree very 
well with the S abundances derived from the \SI\ lines observed with UVES.
An average of \sfe\ = +0.29 is obtained from the UVES lines and 
\sfe\ = +0.27 from the CRIRES lines. For all six \SI\ lines the
mean sulphur abundances is \logS = 5.79 and the rms deviation
is only 0.035\,dex. 
Although one should not put too much weight on a single star,
we consider this good agreement to be an important check of the
reliability of our sulphur abundance determinations.

According to Takeda et al. (\cite{takeda05}), the non-LTE corrections 
for the 1.046\,$\mu$m \SI\ triplet are somewhat smaller than in the
case of the $\lambda 9212.9, 9237.5$\,\AA\  lines. Hence, one
might in principle also use the 1.046\,$\mu$m triplet to study non-LTE 
effects and calibrate the $S_{\rm H}$ parameter like we did when comparing
S abundances from the $\lambda 8694.6$ line and the 
$\lambda\lambda 9212.9, 9237.5$
pair. This would, however, require IR observations for many more stars
to provide statistically significant results.  

\subsection{Zinc}
\label{zinc}
The Zn abundances are derived from the equivalent widths of the two \ZnI\ lines at 
4722.2 and 4810.5\,\AA . Adopted $gf$ values (see Table \ref{tab:A.1})
are from Bi\'{e}mont \& Godefroid (\cite{biemont80}), and
collisional broadening data were taken from Barklem et al. (\cite{barklem00b}).
In Paper I, a solar Zn abundance of \logZnsun = 4.57 was
derived from the two lines, but this value is sensitive to 
the solar microturbulence parameter. Here, we adopt \logZnsun = 4.60
as derived by Bi\'{e}mont \& Godefroid (\cite{biemont80})
from six \ZnI\ lines. This value is in good agreement with 
the meteoritic value, \logZn\ = $4.61 \pm 0.04$ (Asplund et al.
\cite{asplund05}).

\begin{figure}
\resizebox{\hsize}{!}{\includegraphics{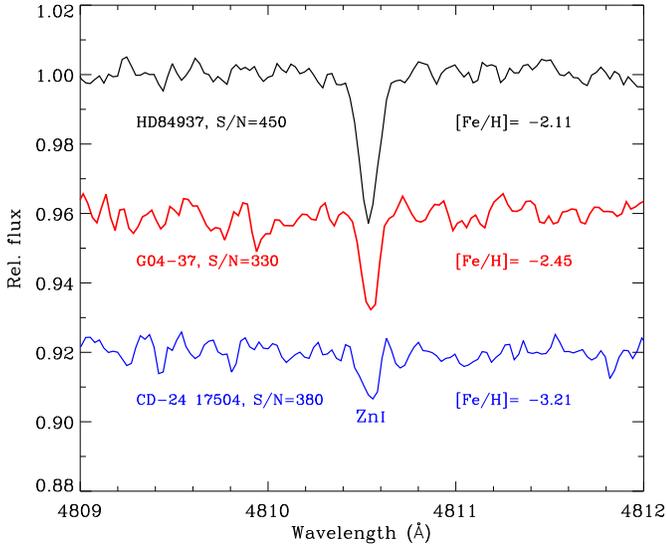}}
\caption{The $\lambda 4810.5$ \ZnI\ line in UVES spectra of three stars.
The spectra of \object{G\,04-37} and \object{CD\,$-24\degr17504$} have
been offset in relative flux by 0.04 and 0.08, respectively.}
\label{fig:zinc-4810}
\end{figure}

As seen from Table \ref{tab:A.1}, the zinc lines are very weak in turnoff stars
with metallicities below  $-2$. The equivalent widths are on the
order of 2--3\,m\AA\ in stars with $\feh \sim -2.5$.
Still, the Zn lines can be clearly detected at metallicities
below $-3$ as seen from Fig. \ref{fig:zinc-4810}. This is connected to
the fact that the Zn/Fe ratio appears to rise steeply below $\feh \sim -3$.

Takeda et al. (\cite{takeda05}) have also
performed non-LTE calculations for the $\lambda\lambda 4722.2,4810.5$ \ZnI\ 
lines. Generally, the absolute value of the non-LTE corrections are smaller 
than in the case of the  $\lambda\lambda 9212.9,9237.5$ \SI\ lines, and 
they have opposite sign. For a typical metal-poor turnoff star, the non-LTE
correction of the zinc abundance is on the order of +0.10 dex, whereas the 
correction is $-0.25$\,dex for the sulphur abundance. For cooler and
more metal-rich stars the non-LTE corrections on zinc are less than 0.1\,dex.
These corrections refer to a hydrogen collisional parameter $S_{\rm H} = 1$,
and increase somewhat if $S_{\rm H} < 1$.

As discussed in Paper I, the absolute values of the 3D corrections of 
the Zn abundances are less than 0.1 dex and they are similar to the 3D corrections
of the Fe abundances based on \FeII\ lines. Hence, the derived Zn/Fe ratio would
not change significantly (at least not under the assumption of LTE)
if 3D models were applied.

\section{Results and discussion}
\label{results}

\subsection{S/Fe vs. Fe/H}
\label{sfe:results}
The derived sulphur abundances are given in Table \ref{table:par}.  
In the calculation of \sfe , solar abundances \logSsun\ = 7.20 and
\logFesun\ = 7.50 were adopted.
Had we instead applied the newest solar abundances, 
\logSsun = 7.14 and \logFesun = 7.45, determined by Asplund et al. 
(\cite{asplundetal05}), the \sfe\ values would have increased by only
0.01\,dex.

\begin{figure}
\resizebox{\hsize}{!}{\includegraphics{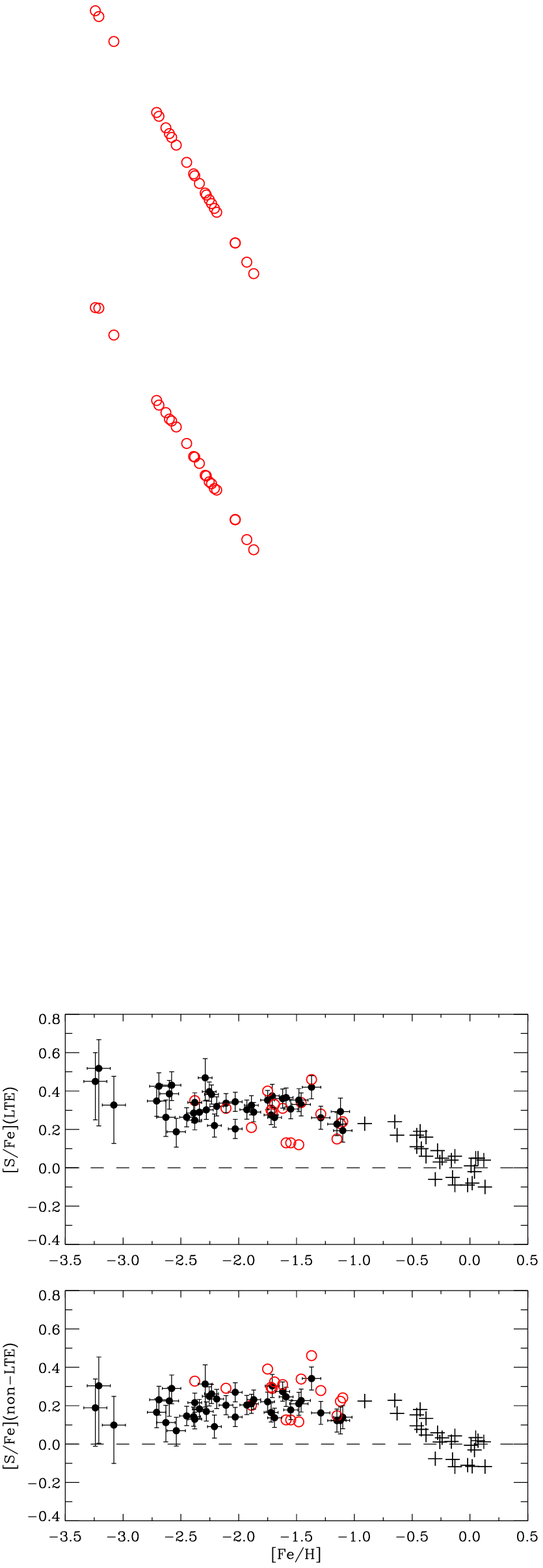}}
\caption{[S/Fe] vs. [Fe/H] for our sample of halo stars 
supplemented with disk stars (shown as crosses) from Chen et al. (\cite{chen02}).
Filled circles with error bars show data based on  S abundances derived from
the $\lambda \lambda 9212.9, 9237.5$ \SI\ lines, and open circles show data
based on the weak $\lambda 8694.6$ \SI\ line. 
In the upper panel LTE has been assumed in deriving the S abundances,
whereas the lower
panel includes non-LTE corrections from Takeda et al. (\cite{takeda05})
with the hydrogen collisional parameter $S_{\rm H}$ = 1.
In both cases, the iron abundances are the LTE values derived from 
\FeII\ lines.}
\label{fig:SFe-FeH}
\end{figure}

The \sfe\ values in Table \ref{table:par} are plotted vs. 
\feh\ in Fig. \ref{fig:SFe-FeH} together with data for 25 disk stars from
Chen et al. (\cite{chen02}). The error bars shown refer to the 
1-$\sigma$ statistical error caused by the uncertainty of 
the observed equivalent widths and the atmospheric parameters   
of the stars. In the case of \feh , which is based on many \FeII\
lines, the dominating error comes from the uncertainty of the gravity;
$\sigma (\logg) = 0.15$\,dex corresponds to $\sigma \feh = 0.05$\,dex
(see Table 4 in Paper\,I). The \logg\ induced error of \sfe\ is,
on the other hand, negligible, because a change in gravity has
nearly the same effect on S and Fe abundances. In the case
of \sfe\ the major error contribution comes from the uncertainty of \teff\ 
and the error of the equivalent width measurements. For the three most
metal-poor stars with $\feh < -3$, the $W_\lambda$ induced error becomes particular
large, because only the $\lambda 9212.9$ \SI\ line is strong enough
to be detected and it is as weak as $W_\lambda \sim 5$\,m\AA .


As seen from Fig. \ref{fig:SFe-FeH}, the LTE values of \sfe\ suggest a small
slope of \sfe\ as a function of \feh , whereas the non-LTE values 
give a nearly flat relation.   
The mean \sfe\ and the rms deviation for the halo stars are:

\smallskip
\noindent
$<\!\sfe \!> \: =  0.319$ $\pm 0.078$ in the LTE case, and

\noindent
$<\!\sfe \! > \: =  0.208$ $\pm 0.067$ in the non-LTE case. 

\smallskip
We note that the four single-lined spectroscopic binaries (see Table
\ref{table:par}) do not show any particular large or systematic deviations
from the mean value of \sfe .


The trend of \sfe\ in Fig. \ref{fig:SFe-FeH} corresponds rather well
to analogous trends 
recently derived for \mgfe , \sife\ and \cafe\ (Jonsell et al. 
\cite{jonsell05}, Gehren et al. \cite{gehren06}). Qualitatively,
our data also agree with the trend of \sfe\ predicted from models 
of the chemical evolution of our Galaxy.
Based on the Woosley \& Weaver (\cite{woosley95}) yields of Type II SNe,
Goswami \& Prantzos (\cite{goswami00}) obtain a near-constant level 
$\sfe \sim +0.4$ in the metallicity range $-3 < \feh < -1$ 
and a decline of \sfe\
for $\feh > -1$ due to the additional supply of iron from Type Ia SNe. 
Chiappini et al. (\cite{chiappini99}), on the other hand, obtain a 
slope of \sfe\ in the range $-3 < \feh < -1$ like the one we determine
in the LTE case, because they assume that Type Ia SNe start contributing
with iron already in the Galactic halo phase. Finally, Kobayashi et al.
(\cite{kobayashi06}) predict a level of $\sfe \sim +0.4$ based on
new yield calculations in a model where hypernovae are assumed to have
the same frequency as Type II SNe. We note that our
data do not allow us to establish conclusively whether a small slope
of \sfe\ is present or not due to
the uncertainty about the non-LTE corrections. It should also be noted
that the predicted level of \sfe\ among halo stars depends
critically on the assumed mass cut between the ejecta and the 
collapsing core in massive SNe. Hence, any detailed comparison
between observed and predicted trends of \sfe\ seems somewhat premature.

It is interesting to compare the observed scatter of
\sfe\ with the scatter predicted from stochastic
chemical evolution models. In sharp contrast to the results of
Caffau et al. (\cite{caffau05}), we find a very small scatter
($\pm 0.07$\,dex) of \sfe\ among the halo stars, and there is no indication
that the scatter increases significantly towards the most metal-poor stars.
This scatter is not much higher than expected from the observational
errors of \sfe . Hence, the cosmic scatter in \sfe\ at a given
\feh\ must be less than 0.07\,dex. A similar low scatter
has been obtained for other $\alpha$-capture elements, when high
precision and  homogeneous data have become available. Nissen et al.
(\cite{nissen94}) found the scatter of Mg, Ca and Ti relative to Fe
to be $<\!0.06$\,dex in the metallicity range $-3 < \feh < -1.5$. More
recently, Arnone et al. (\cite{arnone05}) found the cosmic scatter of
\mgfe\ to be $<\!0.06$\,dex for 25 turnoff stars 
in the range $-3.2 < \feh < -2$\footnote{This 
range is on our metallicity scale. The metallicities 
of Arnone et al. (\cite{arnone05}) are based on \FeI\ lines and a 
\teff\ scale that is on the average $\sim 200$\,K lower than our scale. 
For 14 stars in common with our sample, they get \feh\ values that are 
on the average 0.20\,dex lower than our values.}.  
At still lower metallicities, Cayrel et al. (\cite{cayrel04}) find
a scatter of 0.10 - 0.15\,dex for \mgfe , \sife , \cafe\ and \tife\ for 
30 giant stars ranging in \feh\ from $-4.1$ to $-2.7$.
A similar low scatter of these ratios has been derived by 
Cohen et al. (\cite{cohen04}) for 28 dwarf stars in the metallicity range
$-3.6 < \feh < -2.0$.

The observed scatter of $\alpha$-capture elements 
relative to Fe is much smaller than predicted from stochastic models
of the chemical evolution of metal-poor systems (Argast et al.
\cite{argast00}, \cite{argast02}, Karlsson \& Gustafsson \cite{karlsson05}).
This is connected to the fact that calculated yields of Type II SNe
vary strongly with progenitor mass. 
According to Nomoto et al. (\cite{nomoto97}), the yield ratio S/Fe increases
by a factor of 12 when the progenitor mass is changed from 
13 to 40 solar masses.
Hence, in the early Galaxy, where only a few
supernovae enrich the interstellar gas of a star-forming cloud according to
current models, one would expect a much higher scatter in 
\sfe\ than the derived value of $<0.07$\,dex for our sample of stars.
Sulphur has not been specifically modelled in any of the stochastic models,
but in the case
of Si that has similar yield variations as S, Argast et al.
(\cite{argast00}) predict a rms scatter of $\pm 0.35$\,dex at $\feh = -3.0$,
$\pm 0.25$\,dex at $\feh = -2.5$, and $\pm 0.12$\,dex at $\feh = -2.0$. 
As discussed in detail by 
Arnone et al. (\cite{arnone05}) and Karlsson \& Gustafsson 
(\cite{karlsson05}), possible explanations of this discrepancy
are: $i)$ The calculated yield ratios are wrong, i.e. the released
amount of alpha-elements relative to Fe is nearly independent of
progenitor mass, $ii)$ the IMF is biased to a narrow mass range in
the early Galaxy, and $iii)$ the mixing of SNe ejecta is much more
rapid than assumed in the models, such that a large number of SNe
always contribute to the enrichment of a cloud. The last
possibility means that the mixing time scale of supernova ejecta
is considerably shorter than the cooling time of star-forming gas clouds.

It should be noted that the scatter in the abundance of
$\alpha$-capture elements relative to Fe probably increases towards
the more metal-rich end of the halo. Paper\,I includes
six halo stars with $\feh > -1$, four of which have $\sfe \simeq +0.3$
and two have  $\sfe \simeq 0.0$. This scatter is in accordance with
the results of
Nissen \& Schuster (\cite{nissen97}), who find a similar dichotomy in
\alphafe\  for halo stars with $-1 < \feh < -0.6$.
As suggested in their paper, the explanation of the scatter may be that
the metal-rich $\alpha$-poor stars have formed in the outer regions of the
Galaxy where the star formation has proceeded so slowly that iron from
Type Ia SNe has been incorporated in them. 
Supporting evidence has been obtained by Gratton et al. (\cite{gratton03}),
who find \alphafe\ to be lower and more scattered in stars belonging
to the `accretion' component of the Galaxy than in stars belonging 
to the `dissipative' component. 

\subsection{Zn/Fe vs. Fe/H}
\label{zn:results}
\begin{figure}
\resizebox{\hsize}{!}{\includegraphics{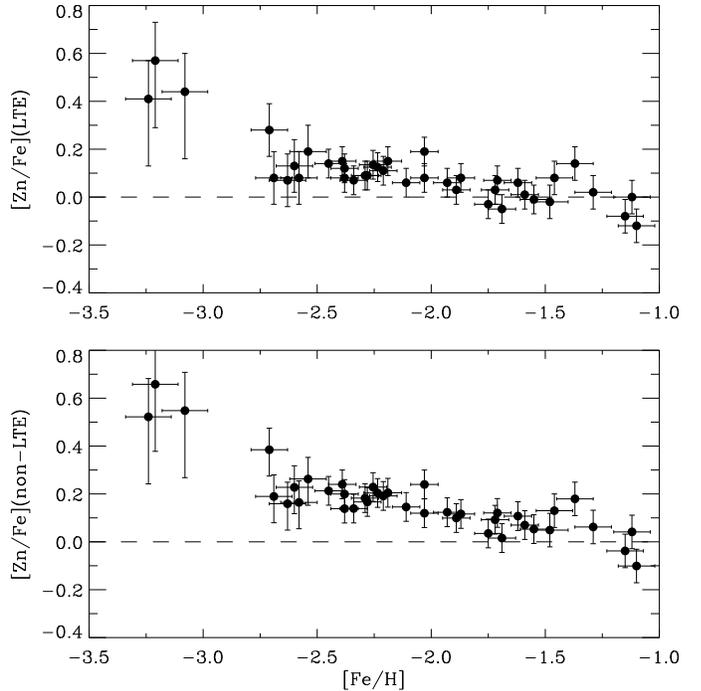}}
\caption{\znfe\ vs. \feh\ for our sample of halo stars.
The upper panel refers to LTE Zn abundances. The lower
panel includes non-LTE corrections from Takeda et al. (\cite{takeda05})
corresponding to a hydrogen collisional parameter $S_{\rm H} = 1$.}
\label{fig:ZnFe-FeH}
\end{figure}

The LTE values of \znfe\ are plotted as a function of
\feh\ in the upper panel of Fig. \ref{fig:ZnFe-FeH}.
In the lower panel, non-LTE corrections of \znfe\ from
Takeda et al. (\cite{takeda05}) have been included. The error bars refer to the
1-$\sigma$ statistical error caused by the uncertainty of
the observed equivalent widths and the atmospheric parameters
of the stars. For the three most
metal-poor stars, the $W_\lambda$-induced error is large,
because the equivalent widths of the \ZnI\ lines are only about 1\,m\AA .

As seen from the upper panel of Fig. \ref{fig:ZnFe-FeH}, the LTE values of
\znfe\ are close to zero in the metallicity range $-2 < \feh < -1$.
In the range $-2.7 < \feh < -2$, all \znfe\ values are, however, 
positive with an average value of $\znfe \simeq +0.1$. Furthermore, \znfe\
seems to rise steeply below $\feh = -3$ to a value of $\znfe \sim +0.5$. 
This trend is reinforced if non-LTE corrections corresponding to $S_{\rm H} = 1$
are applied, because the corrections to \znfe\ are positive and increase
with decreasing \feh . Unlike the case of sulphur, we had no possibility
to calibrate the  $S_{\rm H}$ parameter for zinc (our two \ZnI\ lines
have nearly the same excitation potential), but we note that
if $S_{\rm H} < 1$ (corresponding to a lower efficiency of thermalizing
the Zn atoms by inelastic collisions with hydrogen atoms) then the
non-LTE corrections for the metal-poor stars would be still larger.

The trend of \znfe\ in Fig. \ref{fig:ZnFe-FeH} (LTE case) corresponds
very well to the trend found by Cayrel et al. (\cite{cayrel04}) from an
LTE analysis of the $\lambda\lambda 4722.2,4810.5$ \ZnI\ lines in UVES
spectra of 35 giant stars with $-4.2 < \feh < -2$. Hence, there can be
little doubt that Zn is indeed overabundant with respect to iron for very
metal-poor stars, especially because the possible non-LTE corrections
go in the direction of making the overabundance larger.

Zinc is a key element in recent studies of nucleosynthesis and chemical
evolution in the early Galaxy (Umeda \& Nomoto \cite{umeda03}, \cite{umeda05};
Nomoto et al. \cite{nomoto06}; Kobayashi et al. \cite{kobayashi06}). As shown by  
Kobayashi et al. (\cite{kobayashi06}), traditional yields of Type II SNe
(Nomoto et al. \cite{nomoto97}) correspond to $\znfe \sim -1$, i.e. far
below the observed values of \znfe . In order to explain a level of
\znfe\ around zero, one has to invoke models of hypernovae, i.e. core collapse
SNe with explosion energy $E \simgt 10^{52}$ ergs. Furthermore, a
mixing and fallback mechanism or an asymmetric explosion (Maeda \& Nomoto
\cite{maeda03}) has to be introduced
in order to bring up sufficient Zn from layers with complete Si burning.
By including such hypernovae with a frequency of 50\% relative to Type II SNe
and a Salpeter IMF, Kobayashi et al. (\cite{kobayashi06}) predict 
$\znfe \sim 0.1$ in the Galactic halo. The upturn of \znfe\ at the lowest
metallicities is, however, not predicted. In order to explain $\znfe \simgt +0.3$,
Nomoto et al. (\cite{nomoto06}) suggest that stars with $\feh < -3$
have been formed from the ejecta of Pop. III hypernovae with very 
large explosion energy. Another possibility for high Zn/Fe production is
from core-collapsing, very massive stars with $M \sim 500 - 1000 M_{\odot}$
(Ohkubo et al. \cite{ohkubo06}).

\subsection{S/Zn vs. Zn/H}
\label{szn:results}
As explained in the Introduction, the S/Zn ratio is potentially
of great value in the analysis of the
metallicities of distant galaxies detected via the absorption lines
they produce in the spectra of quasars; the damped Lyman-alpha 
systems (DLAs) are a particularly important subset of such
absorbers (Wolfe at al. \cite{wolfe05}).
The role of S and Zn in these studies stems from the 
fact that both elements have a low condensation temperature
and are not expected to be significantly depleted onto dust.
This is not the case, for example, for refractory elements such
as Si and Fe. 
Indeed, one of the motivations for the present study was to use 
the behaviour of \szn\ vs. \znh\ in Galactic stars as a 
guide to the interpretation of the star formation history
of high redshift galaxies.

\begin{figure}
\resizebox{\hsize}{!}{\includegraphics{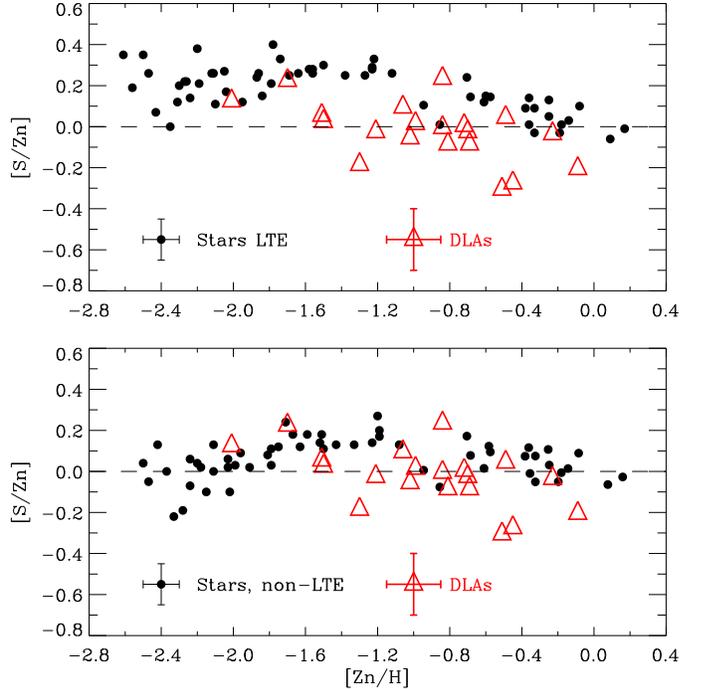}}
\caption{\szn\ vs. \znh\ for Galactic stars and damped Lyman-alpha systems.
Typical 1-$\sigma$ error bars are shown.}
\label{fig:SZn-ZnH}
\end{figure}

Figure \ref{fig:SZn-ZnH} shows \szn\ vs. \znh\ for our halo stars 
except the three
with $\feh < -3$, which have such large errors in both S and Zn that
the ratio S/Zn becomes very uncertain. In addition, we have included six
halo stars with $-1.1 < \znh < -0.6$ from Paper~I and 14 disk stars
with S and Zn abundances from Chen et al. (\cite{chen02}; \cite{chen04}).
The upper panel shows LTE values; in the lower panel non-LTE corrections
($S_{\rm H} = 1$) from Takeda et al. (\cite{takeda05}) have
been applied. 
By comparing the two panels, the importance of the
non-LTE corrections is readily apparent. 
When the corrections are applied, 
there is an overall decrease in \szn\ at all but the highest values 
of \znh\ considered here.
Furthermore, non-LTE effects are most
significant at low metallicities with the result that,
apparently, \szn\ reverts to solar values when
$\znh \simlt -2$. 
Such behaviour is unusual but, given our current limited
understanding of the nucleosynthesis of Zn, cannot be excluded.

Taken at face value, the lack of a strong metallicity trend 
in the lower panel of Fig.~\ref{fig:SZn-ZnH} would indicate
that the usefulness of the S/Zn ratio as a `clock' of the
star-formation history is rather limited---departures from
the solar value are generally small and, in particular,
not much greater than the typical
measurement error in DLAs. 
In  Table C.1 we have collected
all published, or about to be published, 
measurements of \szn\ in DLAs secured with
echelle spectrographs on 8-10\,m class telescopes.
In comparison with the corresponding table in Paper I, 
the number of DLAs with accurate measures of the abundances
of sulphur and zinc has doubled since 2004, from 10 to 20.
These data are shown as triangles in Fig.~\ref{fig:SZn-ZnH}.

Whereas in previous papers (e.g. Paper I) a systematic difference
had been noted between the \szn\ ratio in Galactic 
metal-poor stars and in DLAs of comparable metallicities
(evident in the top panel of Fig.~\ref{fig:SZn-ZnH}), the situation
is considerably less clear-cut if one adopts the non-LTE 
corrections to the abundances of S and Zn. 
If these corrections are appropriate, it would explain why
so few DLAs show enhanced \szn\ values, when other
ratios of alpha-capture to iron-peak elements apparently
do, once dust depletions are accounted for 
(e.g. Prochaska \& Wolfe 2002). 
A minor difference -- which needs to be quantified with larger
datasets -- between Galactic stars and DLAs is the larger
scatter in the \szn\ ratio possibly exhibited by the latter.
It is intriguing, in particular, to find an 
{\it underabundance\/} of S relative to Zn
in some DLAs with relatively high overall metallicities.
If confirmed, the larger dispersion of the ratio in DLAs 
could be an indication that, as one might expect, 
the host galaxies of DLAs experienced a variety 
of different star formation histories, and don't necessarily
follow closely the abundance trends seen in 
a {\it single} galaxy like the Milky Way.

\section{Summary and conclusions}
\label{sect:conclusions}
High $S/N$ UVES spectra of 40 metal-poor halo stars have been used
to derive S, Fe and Zn abundances. For one star we also present
novel observations of the \SI\ triplet at 1.046\,$\mu$m
carried out with the ESO VLT CRIRES spectrograph. These data
confirm the S abundance obtained from other \SI\ lines and
demonstrates the high efficiency of CRIRES in obtaining
high-resolution infrared spectra.

From the results presented in this paper, we conclude that \sfe\ in 
Galactic halo stars shows the same kind of dependence on \feh\ as
\mgfe , \sife\  and \cafe , i.e. a near-constant ratio at a level of
+0.2 to +0.3\,dex. This strongly suggests that sulphur in the Galactic
halo was made by $\alpha$-capture processes in massive SNe, as 
predicted from current models of Galactic chemical evolution with
yields calculated for hydrostatic and explosive oxygen and
silicon burning. 

Among the 40 halo stars observed, we do not find a single case of the
high S/Fe ratios ($\sfe > 0.60$) claimed in some recent papers
(Israelian \& Rebolo \cite{israelian01};Takada-Hidai et al. \cite{takada02};
Caffau et al. \cite{caffau05}). The majority of these high values were based on 
the very weak $\lambda 8694.6$ \SI\ line and, as shown in Sect. \ref{sect:sulphur},
our very high S/N, fringe-corrected spectra do not confirm the claimed 
strength of this line in three metal-poor main-sequence stars. 
A more convincing detection of the $\lambda 8694.6$ \SI\ line
has been obtained by Israelian \& Rebolo (\cite{israelian01})
for two giant stars, HD\,2665 and HD\,2796  with \feh\ = $-2.0$ and $-2.3$. 
The derived \sfe\ values are +0.69 and +0.81 dex, respectively.
Further studies of S abundances in such metal-poor giant stars
would be interesting, especially if one could use
the forbidden sulphur line at 1.082\,$\mu$m, which 
is expected to be insensitive to non-LTE effects. 

From an observational point of view, the 
$\lambda \lambda 9212.9, 9237.5$ pair of \SI\ lines 
or the \SI\ triplet at 1.046\,$\mu$m provide more precise values
of the S abundance than the weak $\lambda 8694.6$ \SI\ line, but 
according to Takeda et al. (\cite{takeda05}) these stronger lines are quite
sensitive to departures from LTE.
The size of the non-LTE effects depends on 
the efficiency of inelastic collisions with neutral hydrogen.
Adopting the classical formula of Drawin (\cite{drawin68}, \cite{drawin69})
with a scaling parameter $S_{\rm H}$, a comparison of S abundances from the 
$\lambda 8694.6$ line and the $\lambda \lambda 9212.9, 9237.5$ pair points
to $S_{\rm H} \simgt 1$, which means that we are not too far away
from LTE. Nevertheless, the  non-LTE corrections may be of importance for the 
derived trend of \sfe\ as seen
from Fig. \ref{fig:SFe-FeH}. Hence, it would be important to
perform improved quantum mechanical
calculations for inelastic S + H collisions as has
been achieved for Li + H collisions (Belyaev \& Barklem \cite{belyaev03}).

An interesting and robust result of our investigation is the
small scatter of \sfe\ at a given metallicity. It is found to be around
$\pm 0.07$\,dex in the whole range $-3.2 < \feh < -1$. A similarly low scatter
has been found for the other $\alpha$-capture elements, i.e. \mgfe ,
\sife , and \cafe\ 
(Cayrel et al. \cite{cayrel04}; Cohen et al. \cite{cohen04}).
Current stochastic models
of the chemical evolution of metal-poor systems (Argast et al.
\cite{argast00}, \cite{argast02}, Karlsson \& Gustafsson \cite{karlsson05})
predict a much higher scatter of \alphafe\ for metallicities below $-2$.
Clearly the models and/or calculated yield ratios of massive SNe 
need to be revised.

Zinc is found to be slightly overabundant with respect to Fe in the 
metallicity range $-3 < \feh < -2$. Below $\feh \sim -3$, \znfe\ increases
rapidly to a level of +0.5\,dex. Hence, our study of main-sequence halo stars
confirms the upturn of Zn/Fe previously found for very metal poor giants
(Primas et al. \cite{primas00}; Johnson \& Bolte \cite{johnson01};
Cayrel et al. \cite{cayrel04}). The high value of Zn/Fe at the lowest
metallicities may be a signature that stars with $\feh < -3$ have been
made from the ejecta of hypernovae with very large explosion energy as
suggested by Nomoto et al. (\cite{nomoto06}). However, 
our understanding of the nucleosynthesis of
Zn at all metallicities is still very limited.

Finally, we find a much reduced `signal' in the \szn\ ratio
as a tracer of the previous history of star formation, if
non-LTE effects to the abundances of these two elements
are taken into account. 
It would be very desirable to perform a check of the 
validity of the non-LTE corrections
to the Zn abundance by comparing lines arising
from levels with different excitation potentials,
as we have been able to do here for S.
However, if the non-LTE corrections 
applied are appropriate, it would explain why
in general DLAs do not exhibit markedly supersolar
\szn\ ratios, even at  metallicities  
comparable to those of Galactic halo stars.

\begin{acknowledgements}
The ESO staff at Paranal is thanked for carrying out the VLT/UVES
service observations. We acknowledge help from Francesca Primas,
Hughes Sana, Lowell Tacconi-Garman and Burkhard Wolff in obtaining 
the CRIRES spectrum of \object{G\,29-23}. Paul Barklem is thanked 
for providing a version of the {\sc bsyn} program including the
hydrogen lines. We are grateful to Miroslava Dessauges-Zavadsky, 
Sara Ellison and Jason Prochaska for communicating measurements of 
[S/Zn] in DLAs in advance of publication.
This publication made use of the SIMBAD database operated
at CDS, Strasbourg, France, and of data products from the Two Micron All
Sky Survey, which is a joint project of the University of Massachusetts and
the Infrared Processing and Analysis Center/California Institute of 
Technology, funded by NASA and the National Science Foundation.  
This research has made use of NASA's Astrophysics Data System.
\end{acknowledgements}

\begin{appendix}
\section{Line list and equivalent widths}
\label{appendix:A}
 
Table \ref{tab:A.1} gives a list of spectral lines used in this
paper and the equivalent widths measured in the 2004 UVES spectra.
Where no value is given,
it is either because the line is too weak to provide
a reliable abundance, or, in the case of the
$\lambda \lambda 9212.9, 9237.5$ \SI\ lines, is affected
by residuals from the removal of strong telluric \water\ lines.
We also note that the \SI\
line at 9228.1\,\AA\ is not included, because it falls close to the
center of the Paschen-zeta line. 

\begin{table*}
\caption{List of spectral lines and equivalent widths measured 
in the UVES 2004 spectra of the following twelve stars:
(1) \object{CD\,$-24\degr17504$},
(2) \object{CD\,$-71\degr1234$},
(3) \object{CS\,22943-095},
(4) \object{G\,04-37},
(5) \object{G\,48-29},
(6) \object{G\,59-27},
(7) \object{G\,126-52},
(8) \object{G\,166-54},
(9) \object{HD\,84937},
(10) \object{HD\,338529},
(11) \object{LP\,635-14},
(12) \object{LP\,651-4}.}
\label{tab:A.1}
\setlength{\tabcolsep}{0.2cm}
\begin{tabular}{lccrrrrrrrrrrrrr}
\noalign{\smallskip}
\hline\hline
\noalign{\smallskip}
      &             &          &         &                        \multicolumn{12}{c}{Equivalent width}                        \\
  ID  &  Wavlength  & Exc.pot  & log$gf$ & (1) &  (2) &  (3) &  (4) &  (5) &  (6) &  (7) &  (8) &  (9) &  (10) &  (11) &  (12) \\
      &    \AA      &   eV     &         &                        \multicolumn{12}{c}{m\AA }                                   \\
\noalign{\smallskip}
\hline
\noalign{\smallskip}
\SI   &  8694.64 &  7.87 &     0.03 &       &       &       &       &       &       &       &       &   1.6 &       &       &       \\
\SI   &  9212.87 &  6.52 &     0.38 &   5.7 &  21.0 &  31.3 &  17.6 &  15.2 &  40.6 &  24.7 &  17.5 &  38.3 &  33.5 &  21.9 &  12.3 \\
\SI   &  9237.54 &  6.52 &     0.01 &       &  10.9 &  14.5 &   6.7 &   7.8 &  21.6 &       &   8.1 &  16.9 &  15.5 &   9.0 &       \\
\noalign{\smallskip}
\FeI  &  4199.10 &  3.05 &     0.16 &  10.3 &  36.9 &  41.8 &  34.7 &  23.4 &  55.2 &  41.3 &  26.2 &  46.9 &  42.1 &  36.3 &  25.4 \\
\FeI  &  4233.60 &  2.48 &  $-$0.58 &   6.5 &  28.4 &  33.2 &  26.1 &  14.4 &  49.4 &  32.9 &  18.2 &  38.7 &  33.0 &  26.6 &  17.2 \\
\FeI  &  4250.12 &  2.47 &  $-$0.38 &   9.8 &  37.8 &  42.6 &  34.4 &  22.4 &  58.7 &  42.2 &  25.3 &  48.3 &  43.2 &  36.3 &  24.9 \\
\FeI  &  4271.15 &  2.45 &  $-$0.34 &  11.5 &  41.3 &  45.6 &  39.0 &  26.0 &  63.6 &  46.0 &  28.8 &  52.7 &  46.6 &  39.6 &  26.9 \\
\FeI  &  4282.40 &  2.18 &  $-$0.78 &   7.5 &  30.2 &  35.7 &  27.9 &  17.7 &  51.7 &  35.3 &  20.1 &  41.5 &  35.7 &  29.4 &  18.4 \\
\FeI  &  4918.99 &  2.86 &  $-$0.34 &   5.5 &  25.3 &  29.9 &  22.8 &  13.0 &  45.9 &  30.1 &  15.4 &  35.1 &  30.0 &  24.0 &  14.9 \\
\FeI  &  4920.50 &  2.83 &     0.07 &  12.8 &  44.6 &  49.7 &  42.4 &  28.9 &  66.9 &  50.5 &  31.5 &  56.1 &  50.3 &  43.0 &  31.2 \\
\noalign{\smallskip}
\FeII &  4178.86 &  2.58 &  $-$2.61 &   1.9 &  11.2 &  14.6 &   9.0 &   6.6 &  25.7 &  14.5 &   7.9 &  19.9 &  15.8 &  11.5 &   6.1 \\
\FeII &  4233.17 &  2.58 &  $-$2.01 &   6.0 &  32.3 &  37.6 &  28.4 &  20.1 &  51.5 &  39.0 &  19.9 &  45.8 &  40.1 &  33.0 &  19.4 \\
\FeII &  4416.83 &  2.78 &  $-$2.65 &       &   7.1 &   9.4 &   6.0 &   4.2 &  16.8 &   9.7 &   4.5 &  13.4 &  10.0 &   7.7 &   4.3 \\
\FeII &  4489.18 &  2.83 &  $-$2.96 &       &   3.3 &   4.6 &   3.0 &       &   9.1 &   4.7 &   2.1 &       &   4.8 &   3.2 &   1.5 \\
\FeII &  4491.41 &  2.85 &  $-$2.80 &       &   4.3 &   6.1 &   3.6 &   2.4 &  11.2 &   6.3 &   2.8 &   8.6 &   6.3 &   4.9 &   2.9 \\
\FeII &  4508.29 &  2.85 &  $-$2.41 &       &  11.1 &  13.8 &   9.2 &   5.4 &  23.7 &  13.9 &   6.7 &  18.6 &  15.0 &   9.7 &   6.2 \\
\FeII &  4515.34 &  2.84 &  $-$2.56 &       &   7.7 &  10.3 &   6.7 &   5.3 &  18.2 &  11.2 &   4.8 &  14.1 &  11.7 &   7.7 &   4.6 \\
\FeII &  4520.23 &  2.81 &  $-$2.66 &       &   7.4 &   9.7 &   5.8 &   3.8 &  16.2 &   9.6 &   4.2 &  12.3 &  10.1 &   7.3 &   3.6 \\
\FeII &  4522.63 &  2.84 &  $-$2.22 &   2.4 &  16.1 &  19.8 &  12.6 &   9.0 &  32.6 &  21.1 &   9.0 &  26.7 &  21.4 &  16.3 &   8.7 \\
\FeII &  4541.52 &  2.85 &  $-$3.04 &       &   3.4 &   3.7 &       &       &   7.2 &   3.7 &   1.5 &   5.6 &   4.6 &   2.7 &   1.5 \\
\FeII &  4555.89 &  2.83 &  $-$2.43 &       &  10.4 &  13.6 &   9.2 &   6.2 &  23.4 &  14.8 &   5.5 &  18.6 &  15.0 &  11.0 &       \\
\FeII &  4576.34 &  2.84 &  $-$3.01 &       &   3.4 &   4.0 &   1.9 &       &   7.3 &   4.1 &   1.7 &   5.7 &   4.5 &   2.6 &   3.0 \\
\FeII &  4583.84 &  2.81 &  $-$1.91 &   4.9 &  27.8 &  33.4 &  23.8 &  17.8 &  47.5 &  34.3 &  17.5 &  41.5 &  35.9 &  29.0 &  17.2 \\
\FeII &  4923.93 &  2.89 &  $-$1.45 &  12.2 &  46.9 &  52.4 &  42.5 &  33.4 &  67.5 &  54.8 &  33.2 &  61.9 &  56.0 &  47.8 &  33.7 \\
\noalign{\smallskip}
\ZnI  &  4722.15 &  4.03 &  $-$0.39 &   0.9 &   2.8 &   2.8 &   2.0 &   1.2 &   5.9 &   3.1 &   1.1 &   3.3 &   3.2 &   2.4 &   0.9 \\
\ZnI  &  4810.53 &  4.08 &  $-$0.17 &   1.4 &   3.0 &   4.9 &   3.1 &   1.7 &   7.5 &   4.2 &   1.9 &   5.5 &   4.3 &   3.3 &   2.0 \\
\hline
\end{tabular}
\end{table*}

\end{appendix}

\begin{appendix}
\section{Photometric and spectroscopic indices}
\label{appendix:B}

In Table \ref{tab:B.1} the magnitudes $V$ and $K_s$ 
for our sample of stars are given together with the Str\"{o}mgren
indices \by , $m_1$, $c_1$ and the photometric index of the H$\beta$ line,
$\beta$(phot). The table also gives the value of the $\beta$(UVES)
index measured from the profile of H$\beta$ in the UVES spectra.
This index is defined as
\begin{eqnarray}
\beta({\rm UVES})  =  10 \cdot \frac{F_{\rm Cb} + F_{\rm Cr}}{F_{\rm Lb} + F_{\rm Lr}},
\end{eqnarray}
where $F_{\rm Cb}$, $F_{\rm Cr}$ are the fluxes in two 8\,\AA\ wide 
pseudo-continuum bands centered $\pm 20$\,\AA\ from the 
H$\beta$ line center  and $F_{\rm Lb}$, $F_{\rm Lr}$ the 
fluxes in two 10\,\AA\ wide line bands  centered $\pm 7$\,\AA\ 
from the center of H$\beta$. As seen from Fig. \ref{fig:H-beta}, the fluxes in the 
line bands are more affected by variations in effective temperature
than the fluxes in the `continuum' bands. Hence, the index becomes
sensitive to \teff .

\begin{figure}
\resizebox{\hsize}{!}{\includegraphics{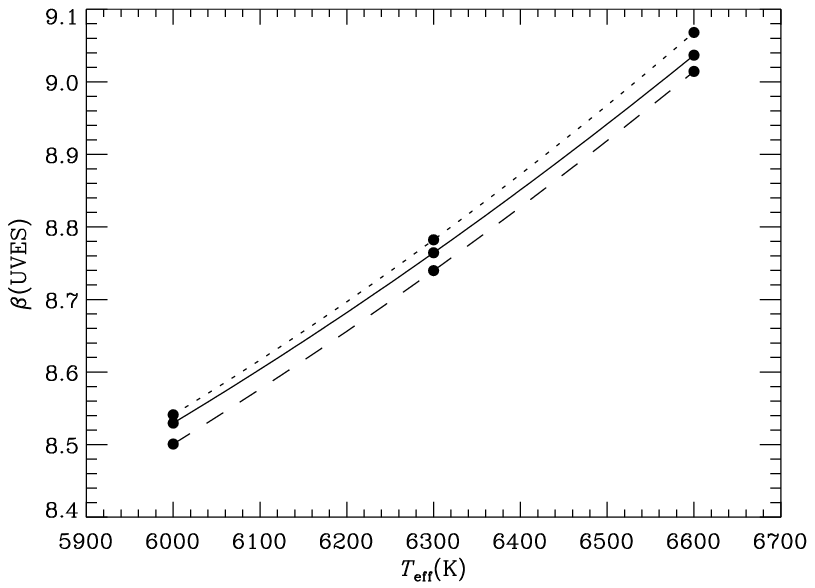}}
\caption{The relation between $\beta$(UVES) and \teff\ as
calculated for a set of model atmospheres. The full drawn line
corresponds to \logg\ = 4.3, \feh\ = $-2.0$, the dotted line
to  \logg\ = 3.9, \feh\ = $-2.0$, and the dashed line to
\logg\ = 4.3, \feh\ = $-3.2$.}
\label{fig:betaUVES-Teff}
\end{figure}

\teff\ was derived from the observed value of $\beta$(UVES)
by quadratic interpolation between theoretical $\beta$(UVES)
values calculated for a grid of 105 model atmospheres with

\smallskip
\noindent
\teff\ = 5400, 5700, 6000, 6300, 6600\,K,

\noindent
\logg\ = 3.5, 3.9, 4.3, and

\noindent
\feh\ = $-0.8, -1.2, -1.6, -2.0, -2.4, -2.8, -3.2$.

\smallskip
In Fig. \ref{fig:betaUVES-Teff} some of the calculated 
$\beta$(UVES) values are plotted as a function of \teff\
together with the interpolation lines. As seen, the variation
of $\beta$(UVES) with \logg\ and \feh\ is small compared 
to the change of $\beta$(UVES) as a function of \teff . For an 
observational error $\pm 0.0 15$ of $\beta$(UVES)
and errors $\pm 0.15$\,dex of \logg\ and \feh , \teff\ is 
determined to a precision of about $\pm 20$\,K.

\begin{figure}
\resizebox{\hsize}{!}{\includegraphics{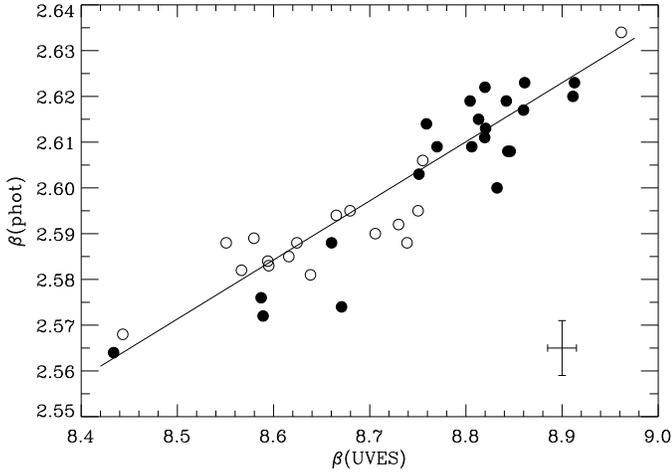}}
\caption{The photometric index of the H$\beta$ line vs. the
index measured from the profile of H$\beta$ in UVES spectra.
1-$\sigma$  error bars: $\sigma (\beta$(UVES))\,=\,$\pm 0.015$ and
$\sigma (\beta$(phot))\,=\,$\pm 0.006$ are shown in the lower right corner.
Stars having $\feh < -2.0$ are plotted with filled circles, and stars having
$-2.0 < \feh < -1.0$ with open circles.}
\label{fig:beta-betaUVES}
\end{figure}

As seen from Fig. \ref{fig:beta-betaUVES} there is a good correlation
between $\beta$(phot) and $\beta$(UVES).
A maximum likelihood fit that takes into account the estimated errors
of the two indices gives the following relation:
\begin{eqnarray}
\beta({\rm phot}) = 1.474 + 0.129 \cdot \beta({\rm UVES})
\end{eqnarray}
with a reduced $\chi ^2$ close to one. The dominant contribution to the
scatter in Fig. \ref{fig:beta-betaUVES} comes from the error of $\beta$(phot).
According to Eq.\,(B.2), the error of $\beta$(UVES) ($\pm 0.015$)
corresponds to an error of $\pm 0.002$ in $\beta$(phot), which is a factor of
three lower than the actual error of the $\beta$(phot) observations
(Schuster \& Nissen \cite{schuster88}).
Hence, we have used Eq.\,(B.2) to predict an improved $\beta$(phot) value
of our stars before calculating the interstellar reddening excess
from the $(b-y)_0$ calibration equations of Schuster \& Nissen 
(\cite{schuster89}) with a zero point shift of 0.005 mag. added
(Nissen \cite{nissen1994}). The so derived value of
$E(b-y)$ is given in column 10 of Table \ref{tab:B.1}.

\begin{figure}
\resizebox{\hsize}{!}{\includegraphics{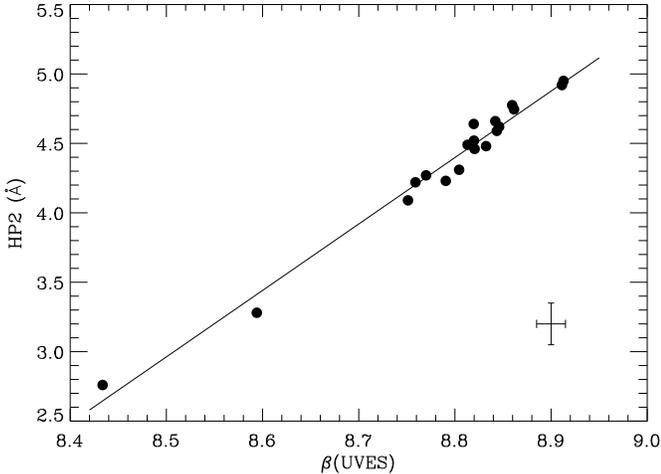}}
\caption{The HP2 index of the H$\delta$ line (from Ryan et al.
\cite{ryan99}) vs. $\beta$(UVES).}
\label{fig:HP2-betaUVES}
\end{figure}

From this discussion it also follows that the error of \teff\ 
determined from the photometric H$\beta$ index would be three
times higher than the error of \teff\ determined from $\beta$(UVES).
On the other hand, it is seen from Fig.
\ref{fig:HP2-betaUVES} that the HP2 index
of the H$\delta$ line (Beers et al. \cite{beers99})
correlates extremely well with $\beta$(UVES) for 19 stars in common with 
Ryan et al. (\cite{ryan99}). Hence, for the group of metal-poor turnoff
stars, one may use the HP2 index, which is based on medium-resolution
spectra, to determine differential values of \teff\ with the same high
precision as obtained from $\beta$(UVES).

The absolute visual magnitude of a star, $M_{V}$(phot), is determined from 
the Str\"{o}mgren indices \by , $m_1$, $c_1$ using a calibration 
derived by Schuster et al. (\cite{schuster06}) for a sample of stars
with accurate ($\sigma (\pi) / \pi < 0.15$)  Hipparcos parallaxes. 
Reddening corrections were
applied if $E(b-y) >0$. The derived values of $M_{V}$(phot) are
given in column 11 of Table \ref{tab:B.1}. In columns 12 and 13, the
absolute magnitude derived directly from the Hipparcos parallax
and its error are given for stars with a parallax error less than 30\%.

\begin{table*}
\caption{$V$ and $K_s$ magnitudes,  Str\"{o}mgren indices, and
photometric and spectroscopic indices of the strength of the H$\beta$ line.
The derived interstellar reddening of \by\ is given in column 10, and the absolute
visual magnitude derived from the Str\"{o}mgren photometry is given in col. 11. 
Columns 12 and 13 give the absolute visual magnitude and its 1-$\sigma$ error derived from the 
Hipparcos parallax if available with an error less than 30\%.}
\label{tab:B.1}
\setlength{\tabcolsep}{0.15cm}
\begin{tabular}{lrrccccccrccc}
\noalign{\smallskip}
\hline\hline
\noalign{\smallskip}
  ID   &   $V$ & $K_s$ &  $b-y$ & $m_1$  &$c_1$ & $\beta$(phot) & Ref.$^{\rm a)}$ & $\beta$(UVES) & $E(b-y)$ & $M_{V}$(phot)
  & $M_{V}$(par) & $\sigma$   \\
\noalign{\smallskip}
\hline
\noalign{\smallskip}
\noalign{\smallskip}
\object{BD\,$-13\degr3442$} & 10.288 &  9.018 & 0.308 & 0.050 & 0.385 & 2.622 & 4 &  8.820 &  0.022 &  3.67 &       & \\      
\object{CD\,$-30\degr18140$} &  9.949 &  8.655 & 0.323 & 0.047 & 0.344 & 2.606 & 1 &  8.755 &  0.019 &  3.88 &  4.19 &  0.46 \\
\object{CD\,$-35\degr14849$} & 10.568 &  9.293 & 0.321 & 0.040 & 0.293 & 2.603 & 1 &  8.751 &  0.011 &  4.47 &       & \\     
\object{CD\,$-42\degr14278$} & 10.216 &  8.770 & 0.361 & 0.040 & 0.229 & 2.576 & 1 &  8.587 &  0.015 &  4.94 &       & \\      
\object{G\,11-44} & 11.091 &  9.740 & 0.335 & 0.061 & 0.248 & 2.588 & 1 &  8.660 & $-$0.005 &  4.80 &       &  \\      
\object{G\,13-09} &  9.998 &  8.739 & 0.311 & 0.048 & 0.373 & 2.609 & 1 &  8.806 &  0.021 &  3.75 &  3.70 &  0.59  \\
\object{G\,18-39} & 10.392 &  9.017 & 0.346 & 0.073 & 0.286 & 2.581 & 1 &  8.638 &  0.002 &  4.35 &       &        \\
\object{G\,20-08} &  9.948 &  8.498 & 0.356 & 0.047 & 0.251 & 2.574 & 1 &  8.671 &  0.020 &  4.79 &  4.47 &  0.43  \\
\object{G\,24-03} & 10.467 &  9.020 & 0.363 & 0.057 & 0.271 & 2.585 & 1 &  8.616 &  0.019 &  4.53 &       &        \\      
\object{G\,29-23} & 10.230 &  8.831 & 0.339 & 0.059 & 0.332 & 2.590 & 1 &  8.706 &  0.017 &  3.87 &       &        \\
\object{G\,53-41} & 11.022 &  9.595 & 0.356 & 0.083 & 0.271 & 2.589 & 1 &  8.580 & $-$0.003 &  4.54 &     &        \\     
\object{G\,64-12} & 11.459 & 10.208 & 0.307 & 0.043 & 0.337 & 2.617 & 4 &  8.860 &  0.020 &  4.37 &       &        \\
\object{G\,64-37} & 11.144 &  9.923 & 0.300 & 0.054 & 0.333 & 2.623 & 4 &  8.861 &  0.007 &  4.31 &       &        \\
\object{G\,66-30} & 11.028 &  9.789 & 0.305 & 0.072 & 0.358 & 2.634 & 1 &  8.962 &  0.016 &  4.20 &       &        \\
\object{G\,126-62} &  9.478 &  8.075 & 0.330 & 0.063 & 0.327 & 2.588 & 1 &  8.739 &  0.012 &  4.00 &  4.06 &  0.37 \\
\object{G\,186-26} & 10.829 &  9.590 & 0.306 & 0.041 & 0.339 & 2.608 & 1 &  8.844 &  0.019 &  4.31 &  5.12 &  0.50 \\
\object{HD\,106038} & 10.179 &  8.761 & 0.342 & 0.092 & 0.264 & 2.583 & 1 &  8.595 & $-$0.017 &  4.69 &  4.99 &  0.36 \\
\object{HD\,108177} &  9.671 &  8.354 & 0.330 & 0.059 & 0.287 & 2.594 & 1 &  8.665 & $-$0.002 &  4.36 &  4.87 &  0.26 \\
\object{HD\,110621} &  9.906 &  8.566 & 0.337 & 0.067 & 0.324 & 2.595 & 1 &  8.679 &  0.007 &  3.89 &  4.12 &  0.46   \\
\object{HD\,140283} &  7.213 &  5.588 & 0.380 & 0.033 & 0.284 & 2.564 & 1 &  8.434 &  0.013 &  3.83 &  3.36 &  0.12   \\
\object{HD\,160617} &  8.733 &  7.311 & 0.347 & 0.051 & 0.331 & 2.584 & 1 &  8.594 &  0.010 &  3.51 &  3.37 &  0.31   \\
\object{HD\,179626} &  9.210 &  7.680 & 0.373 & 0.095 & 0.293 & 2.588 & 1 &  8.551 &  0.005 &  4.29 &  3.57 &  0.39   \\
\object{HD\,181743} &  9.687 &  8.274 & 0.351 & 0.052 & 0.224 & 2.582 & 1 &  8.567 & $-$0.001 &  4.97 &  4.95 &  0.34 \\
\object{HD\,188031} & 10.148 &  8.857 & 0.328 & 0.058 & 0.310 & 2.592 & 1 &  8.730 &  0.009 &  4.19 &       &         \\
\object{HD\,193901} &  8.660 &  7.144 & 0.383 & 0.099 & 0.221 & 2.568 & 1 &  8.443 & $-$0.006 &  5.16 &  5.46 &  0.12 \\
\object{HD\,194598} &  8.354 &  6.982 & 0.344 & 0.091 & 0.269 & 2.588 & 1 &  8.624 & $-$0.011 &  4.63 &  4.62 &  0.15 \\
\object{HD\,215801} & 10.044 &  8.642 & 0.334 & 0.049 & 0.330 & 2.572 & 1 &  8.589 & $-$0.001 &  3.47 &       &         \\
\object{LP\,815-43} & 10.912 &  9.650 & 0.304 & 0.048 & 0.382 & 2.623 & 1 &  8.913 &  0.031 &  4.18 &       &         \\
\object{CD\,$-24\degr17504$} & 12.124 & 10.807 & 0.322 & 0.043 & 0.283 & 2.609 & 4 &  8.770 &  0.010 &  4.60 &       & \\     
\object{CD\,$-71\degr1234$} & 10.44  &  9.114 &       &       &      &       & 5 &  8.790 &  0.019 &  3.67 &       &  \\    
\object{CS\,22943-095} & 11.762 & 10.469 & 0.324 & 0.045 & 0.335 & 2.619 & 3 &  8.804 &  0.025 &  4.17 &       &     \\ 
\object{G\,04-37} & 11.426 &  9.974 & 0.363 & 0.031 & 0.306 & 2.614 & 4 &  8.759 &  0.053 &  4.43 &       &      \\
\object{G\,48-29} & 10.467 &  9.264 & 0.298 & 0.057 & 0.351 & 2.620 & 4 &  8.911 &  0.013 &  4.29 &       &      \\
\object{G\,59-27} & 10.892 &  9.560 & 0.321 & 0.058 & 0.301 & 2.595 & 1 &  8.750 &  0.004 &  4.32 &       &      \\
\object{G\,126-52} & 10.996 &  9.719 & 0.322 & 0.045 & 0.347 & 2.608 & 2 &  8.846 &  0.031 &  4.20 &       &      \\
\object{G\,166-54} & 11.005 &  9.731 & 0.324 & 0.043 & 0.322 & 2.619 & 4 &  8.842 &  0.027 &  4.42 &       &      \\
\object{HD\,84937} &  8.332 &  7.062 & 0.303 & 0.056 & 0.354 & 2.613 & 1 &  8.821 &  0.009 &  3.95 &  3.77 &  0.19 \\
\object{HD\,338529} &  9.370 &  8.144 & 0.308 & 0.045 & 0.366 & 2.600 & 1 &  8.832 & 0.023 &  3.98 &  3.48 &  0.46 \\
\object{LP\,635-14} & 11.362 &  9.997 & 0.347 & 0.026 & 0.366 & 2.611 & 4 &  8.820 & 0.060 &  4.02 &       &      \\
\object{LP\,651-4} & 12.053 & 10.842 & 0.321 & 0.023 & 0.340 & 2.615 & 4 &  8.813 &  0.033 &  4.27 &       &      \\
\hline
\end{tabular}
\begin{list}{}{}
\item[$^{\rm a)}$] References of $V$, $b-y$, $m_1$, $c_1$ and $\beta$(phot): (1) Schuster \& Nissen (\cite{schuster88}),
(2) Schuster et al. (\cite{schuster93}), (3) Schuster et al. (\cite{schuster96}), (4) Schuster et al. (\cite{schuster06}),
(5) Ryan (\cite{ryan89}). For all stars, $K_s$ has been taken from the 2MASS catalogue (Skrutski et al.
\cite{skrutskie06}).
\end{list}
\end{table*}

\end{appendix}

\begin{appendix}
\section{S and Zn abundances in DLAs}
\label{appendix:C}

Table \ref{tab:C.1} provides a compilation of S and Zn abundances in 20
DLA systems as derived from high-resolution spectra obtained with 8-10\,m
class telescopes. References to the original works are
given in the last column of the Table. S abundances have been determined
from the $\lambda\lambda 1250, 1253, 1259$ \SII\ triplet and Zn abundances
from the $\lambda \lambda 2026, 2062$ \ZnII\ doublet.
Both \SII\ and \ZnII\ are the major ionization stages of their respective elements
in \HI\ regions, and corrections for unobserved ion stages
are expected to be unimportant (e.g. Vladilo et al. \cite{vladilo01}).
Neither S nor Zn show much affinity for dust
and the problem is further lessened in DLAs which generally
show only mild depletions of even the refractory elements
(Pettini et al. \cite{pettini97}). Hence, the values given in Table~\ref{tab:C.1}
should reflect the true interstellar abundances of
these two elements in the high redshift galaxies giving rise to
the damped Lyman-alpha systems.

\begin{table*}
\caption{S and Zn abundance measurements in DLAs}
\label{tab:C.1}
\setlength{\tabcolsep}{0.30cm}
\begin{tabular}{lrcccccr}
\noalign{\smallskip}
\hline\hline
\noalign{\smallskip}
  \multicolumn{1}{c}{QSO}
& \multicolumn{1}{c}{$z_{\rm abs}$} 
& \multicolumn{1}{c}{log $N$\/(H~I)}
& \multicolumn{1}{c}{log $N$\/(S~II)}
& \multicolumn{1}{c}{log $N$\/(Zn~II)}
& \multicolumn{1}{c}{[Zn/H]$^{\rm a}$}
& \multicolumn{1}{c}{[S/Zn]$^{\rm b}$}
& ~~Ref.$^{\rm c}$\\
    & 
& \multicolumn{1}{c}{(cm$^{-2}$)}
& \multicolumn{1}{c}{(cm$^{-2}$)}
& \multicolumn{1}{c}{(cm$^{-2}$)}
& 
&
& \\ 
\noalign{\smallskip}
\hline
\noalign{\smallskip}
Q0000$-$2620        & 3.3901   & $21.41 \pm 0.08$ & $14.70 \pm 0.03$  & $12.01 \pm 0.05$ & $-2.01$ &   $+0.14$ &   10 \\
Q0013$-$004         & 1.97296  & $20.83 \pm 0.05$ & $15.28 \pm 0.03$  & $12.74 \pm 0.04$ & $-0.70$ &   $-0.01$ &   11 \\
Q0100$+$130         & 2.30903  & $21.37 \pm 0.08$ & $15.09 \pm 0.06$  & $12.47 \pm 0.01$ & $-1.51$ &   $+0.07$ &   2  \\
Q0201$+$365         & 2.4628   & $20.38         $ & $15.29 \pm 0.01$  & $12.76 \pm 0.05$ & $-0.23$ &   $-0.02$ &   14, 12 \\  
Q0407$-$4410        & 2.5505   & $21.13 \pm 0.10$ & $14.82 \pm 0.06$  & $12.44 \pm 0.05$ & $-1.30$ &   $-0.17$ &   8 \\
Q0407$-$4410        & 2.5950   & $21.09 \pm 0.10$ & $15.19 \pm 0.05$  & $12.68 \pm 0.02$ & $-1.02$ &   $-0.04$ &   8 \\
B0528$-$250         & 2.8120   & $21.11 \pm 0.04$ & $15.56 \pm 0.02$  & $13.27 \pm 0.03$ & $-0.45$ &   $-0.26$ &   1  \\
Q0551$-$366         & 1.96221  & $20.50 \pm 0.08$ & $15.38 \pm 0.11$  & $13.02 \pm 0.05$ & $-0.09$ &   $-0.19$ &   6 \\
FJ081240.6$+$320808 & 2.6263   & $21.35         $ & $15.63 \pm 0.08$  & $13.15 \pm 0.02$ & $-0.81$ &   $-0.07$ &   13  \\
0841$+$1256         & 2.37452  & $20.99 \pm 0.08$ & $14.69 \pm 0.05$  & $12.10 \pm 0.02$ & $-1.50$ &   $+0.04$ &   3  \\
0841$+$1256         & 2.47621  & $20.78 \pm 0.08$ & $14.48 \pm 0.10$  & $11.69 \pm 0.10$ & $-1.70$ &   $+0.24$ &   3 \\
SDSS1116$+$4118A    & 2.9422   & $20.28 \pm 0.05$ & $15.01 \pm 0.10$  & $12.40 \pm 0.33$ & $-0.49$ &   $+0.06$ &   5 \\
LBQS~1210$+$1731    & 1.89177  & $20.63 \pm 0.08$ & $14.96 \pm 0.02$  & $12.40 \pm 0.05$ & $-0.84$ &   $+0.01$ &   3 \\
Q1331$+$170         & 1.77637  & $21.14 \pm 0.08$ & $15.08 \pm 0.11$  & $12.54 \pm 0.02$ & $-1.21$ &   $-0.01$ &   2  \\
HE2243$-$6031       & 2.33000  & $20.67 \pm 0.02$ & $14.88 \pm 0.01$  & $12.22 \pm 0.03$ & $-1.06$ &   $+0.11$ &   7 \\
B2314$-$409         & 1.8573   & $20.90 \pm 0.10$ & $15.10 \pm 0.05$  & $12.52 \pm 0.03$ & $-0.99$ &   $+0.03$ &   4 \\
LBQS~2230$+$0232    & 1.86359  & $20.83 \pm 0.05$ & $15.29 \pm 0.06$  & $12.72 \pm 0.05$ & $-0.72$ &   $+0.02$ &   3  \\
Q2231$-$002         & 2.06616  & $20.53 \pm 0.08$ & $15.10 \pm 0.15$  & $12.30 \pm 0.05$ & $-0.84$ &   $+0.25$ &   2  \\
Q2343$+$125         & 2.4313   & $20.35 \pm 0.05$ & $14.71 \pm 0.08$  & $12.45 \pm 0.06$ & $-0.51$ &   $-0.29$ &   9 \\
Q2343--BX415        & 2.5739   & $20.98 \pm 0.05$ & $15.38 \pm 0.03$  & $12.90 \pm 0.06$ & $-0.69$ &   $-0.07$ &   15 \\
\hline
\end{tabular}

\begin{list}{}{}
\item[$^{\rm a}$]
[$\log$ (Zn/H)$_{\rm DLA} - \log$ (Zn/H)$_{\odot}$],
where $\log$ (Zn/H)$_{\odot} = -7.39$ (Asplund et al. 2005).
\end{list}

\begin{list}{}{}
\item[$^{\rm b}$]
[$\log$ (S/Zn)$_{\rm DLA} - \log$ (S/Zn)$_{\odot}$],
where $\log$ (S/Zn)$_{\odot} = 2.55$ (Asplund et al. 2005).
\end{list}

\begin{list}{}{}
\item[$^{\rm c}$]
References---
1: Cent\'{u}rion et al. (\cite{centurion03});
2: Dessauges-Zavadsky et al. (\cite{dessauges04}):
3: Dessauges-Zavadsky et al. (\cite{dessauges06},\cite{dessauges07});
4: Ellison \& Lopez (\cite{ellison01});
5: Ellison et al. (\cite{ellison01});
6: Ledoux et al. (\cite{ledoux02});
7: Lopez et al. (\cite{lopez02});
8: Lopez \& Ellison (\cite{lopez03});
9: Lu et al. (\cite{lu98});
10: Molaro et al. (\cite{molaro00}); 
11: Petitjean et al. (\cite{petitjean02});
12: Prochaska et al. (\cite{prochaska07});
13: Prochaska et al. (\cite{prochaska03});
14: Prochaska \& Wolfe (\cite{prochaska99});
15: Rix et al. (\cite{rix07}).
\end{list}

\end{table*}

\end{appendix}


\begin{thebibliography}{}

\bibitem[1994]{alonso94}
Alonso, A., Arribas, S., \& Mart\'{\i}nez-Roger, C. 1994, A\&AS, 107, 365

\bibitem[1995]{alonso95}
Alonso, A., Arribas, S., \& Mart\'{\i}nez-Roger, C. 1995, A\&A, 297, 197

\bibitem[1996]{alonso96}
Alonso, A., Arribas, S., \& Mart\'{\i}nez-Roger, C. 1996, A\&A, 313, 873

\bibitem[2000]{argast00}
Argast, D., Samland, M., Gerhard, O.E., \& Thielemann, F.\,-K. 2000,
A\&A, 356, 873

\bibitem[2002]{argast02}
Argast, D., Samland, M., Thielemann, F.\,-K., \& Gerhard, O.E. 2002,
A\&A, 388, 842

\bibitem[2005]{arnone05}
Arnone, E., Ryan, S.G., Argast, D., Norris, J.E., \& Beers, T.C. 2005,
A\&A, 430, 507

\bibitem[2005]{asplund05}
Asplund, M. 2005, ARA\&A, 43, 481

\bibitem[2005]{asplundetal05}
Asplund, M., Grevesse, N., \& Sauval, A.J. 2005,
in: Cosmic abundances as records of stellar evolution and nucleosynthesis, 
eds T.G. Barnes III and F.N. Bash,
ASP Conf. series vol. 336, p. 25

\bibitem[1997]{asplund97}
Asplund, M., Gustafsson, B., Kiselman, D., \& Eriksson, K. 1997, A\&A, 318, 521

\bibitem[2006]{asplund06}
Asplund, M., Lambert, D.L., Nissen, P.E., Primas, F., \& Smith, V.V. 2006,
ApJ, 644, 229

\bibitem[2007]{barklem07}
Barklem, P.S. 2007, A\&A, 466, 327 

\bibitem[2005]{barklem05}
Barklem, P.S., \& Aspelund-Johansson, J. 2005, A\&A, 435, 373

\bibitem[2000a]{barklem00a}
Barklem, P.S., Piskunov, N., \& O'Mara, B.J. 2000a, A\&A, 363, 1091

\bibitem[2000b]{barklem00b}
Barklem, P.S., Piskunov, N., \& O'Mara, B.J. 2000b, A\&AS, 142, 467

\bibitem[2002]{barklem02}
Barklem, P.S., Stempels, H.C., Allende Prieto, C., et al. 2002, A\&A, 385, 951

\bibitem[2003]{bagnulo03}
Bagnulo, S., Jehin, E., Ledoux, C., et al. 2003, ESO Messenger, 114, 10

\bibitem[1999]{beers99}
Beers, T.C., Rossi, S., Norris, J.E., Ryan, S.G., \& Shefler, T. 1999, AJ, 117, 981

\bibitem[2003]{belyaev03}
Belyaev, A.K., \& Barklem, P.S. 2003, PhRvA, 68, 062703

\bibitem[1980]{biemont80}
Bi\'{e}mont, E., \& Godefroid, M. 1980, A\&A, 84, 361

\bibitem[1993]{biemont93}
Bi\'{e}mont, E., Quinet, P., \& Zeippen, C.J. 1993, A\&AS, 102, 435

\bibitem[1967]{bridges67}
Bridges, J.M., \& Wiese, W.L. 1967, Phys. Rev., 159, 31

\bibitem[2005]{caffau05}
Caffau, E., Bonifacio, P., Faraggiana, R., et al. 2005, A\&A, 441, 533

\bibitem[2001]{carney01}
Carney, B.W., Latham, D.W., Laird, J.B., Grant, C.E., \& Morse, J.A.
2001, AJ, 122, 3419

\bibitem[2004]{cayrel04}
Cayrel, R., Depagne, E., Spite, M. et al. 2004, A\&A, 416, 1117

\bibitem[2003]{centurion03} Centuri{\'o}n, M., Molaro, P., 
Vladilo, G., et al. 2003, \aap, 403, 55

\bibitem[2002]{chen02}
Chen, Y.Q., Nissen, P.E., Zhao, G., \& Asplund, M. 2002, A\&A, 390, 225

\bibitem[2004]{chen04}
Chen, Y.Q., Nissen, P.E., \& Zhao, G. 2004, A\&A, 425, 697

\bibitem[1999]{chiappini99}
Chiappini, C., Matteucci, F., Beers, T.C., \& Nomoto, K. 1999, ApJ, 515, 226

\bibitem[2004]{cohen04}
Cohen, J.G., Christlieb, N., McWilliam, A., et al. 2004, ApJ, 612, 1107

\bibitem[2006]{cutri06}
Cutri, R.M., Skrutskie, M.F., Van Dyk, S., et al. 2003, Explanatory
Supplement to the 2MASS All Sky Data Release, (IPAC/California Institute
of Technology); http://www.ipac.caltech.edu/2mass

\bibitem[2000]{dekker00}
Dekker, H., D'Odorico, S., Kaufer, A., Delabre, B., \& Kotzlowski, H. 
2000, Proc. SPIE, 4008, 534

\bibitem[2004]{dessauges04}
Dessauges-Zavadsky, M., Calura, F., Prochaska, J.~X., D'Odorico, S., \&
Matteucci, F.\ 2004, \aap, 416, 79

\bibitem[2006]{dessauges06}
Dessauges-Zavadsky, M., Prochaska, J.~X., D'Odorico, S., Calura, F., \&
Matteucci, F.\ 2006, \aap, 445, 93

\bibitem[2007]{dessauges07}
Dessauges-Zavadsky, M., Prochaska, J.~X., D'Odorico, S., Calura, F., \&
Matteucci, F.\ 2007, \aap, submitted

\bibitem[1968]{drawin68}
Drawin, H.W. 1968, Z. Physik, 211, 404

\bibitem[1969]{drawin69}
Drawin, H.W. 1969, Z. Physik, 225, 483

\bibitem[2007]{ellison07}
Ellison, S.~L., Hennawi, J.F., Martin, C.L., \& Sommer-Larsen, J. 2007,
MNRAS, in press (arXiv:0704.1816)

\bibitem[2001]{ellison01}
Ellison, S.L., \& Lopez, S. 2001, \aap, 380, 117

\bibitem[1997]{esa97}
ESA 1997, The Hipparcos and Tycho Catalogues, ESA SP-1200

\bibitem[1982]{elias82}
Elias, J.H., Frogel, J.A., Matthews, K., \& Neugebauer, G. 1982, AJ, 87, 1029

\bibitem[1993]{fuhrmann93}
Fuhrmann, K., Axer, M., \& Gehren, T. 1993, A\&A, 271, 451

\bibitem[2006]{gehren06}
Gehren, T., Shi, J.R., Zhang, H.W., Zhao, G., \& Korn, A.J. 2006,
A\&A, 451, 1065

\bibitem[2003]{gratton03}
Gratton, R.G., Caretta, E., Desidera, S. et al. 2003, A\&A, 406, 131

\bibitem[2000]{goswami00}
Goswami, A., \& Prantzos, N. 2000, A\&A, 359, 191

\bibitem[1965]{henyey65}
Henyey, L., Vardya, M.S, \& Bodenheimer, P. 1965, ApJ, 142, 841

\bibitem[2001]{israelian01}
Israelian, G., \& Rebolo, R. 2001, ApJ, 557, L43

\bibitem[2005]{jonsell05}
Jonsell, K., Edvardsson, B., Gustafsson, B. et al. 2005, A\&A, 440, 321

\bibitem[1966]{johnson66}
Johnson, H.L. 1966, ARA\&A, 4, 193

\bibitem[2001]{johnson01}
Johnson, J.A., \& Bolte, M. 2001, Nucl. Phys. A, 688, 41c

\bibitem[2005]{karlsson05}
Karlsson, T., \& Gustafsson, B. 2005, A\&A, 436, 879

\bibitem[2004]{kaufl04}
K\"{a}ufl, H.U., Ballester, P., Biereichel, P., et al. 2004, Proc. SPIE, 5492, 1218

\bibitem[2006]{kerber06}
Kerber, F., Nave, G., Sansonetti, C.J., et al.  
2006, in: Ground-based and Airborne Instrumentation for
Astronomy, eds  I.~S. McLean and M. Iye. Proceedings of the
SPIE, 6269, pp. 2O

\bibitem[2006]{kobayashi06}
Kobayashi, C., Umeda, H., Nomoto, K., Tominaga, N., \& Ohkubo, T. 2006,
ApJ, 653, 1145

\bibitem[2005]{korn05}
Korn, A.J., \& Ryde, N. 2005, A\&A, 443, 1029

\bibitem[1978]{lambert78}
Lambert, D.L., \& Luck, R.E. 1978, MNRAS, 183, 79

\bibitem[2002]{latham02}
Latham, D.W., Stefanik, R.P., Torres, G., et al. 2002, AJ, 124, 1144

\bibitem[2002]{ledoux02} Ledoux, C., Srianand, R.,
\& Petitjean, P.\ 2002, \aap, 392, 781

\bibitem[2003]{lopez03} Lopez, S., \&
Ellison, S.~L.\ 2003, \aap, 403, 573

\bibitem[2002]{lopez02} Lopez, S., Reimers, D.,
D'Odorico, S., \& Prochaska, J.~X.\ 2002, \aap, 385, 778

\bibitem[1998]{lu98} Lu, L., Sargent, W.~L.~W.,
\& Barlow, T.~A.\ 1998, \aj, 115, 55

\bibitem[1999]{ludwig99}
Ludwig, H.\,-G., Freytag, B., \& Steffen, M. 1999, A\&A, 346, 111

\bibitem[2006]{masana06}
Masana, E., Jordi, C., \& Ribas, I. 2006, A\&A, 450, 735

\bibitem[2003]{maeda03}
Maeda, K., \& Nomoto, K. 2003, ApJ, 598, 1163

\bibitem[2006]{melendez06}
Melend\'{e}z, J., Shchukina, N.G., Vasiljeva, I.E., \& Ram\'{\i}rez, I.
2006, ApJ, 642, 1082

\bibitem[2000]{molaro00} Molaro, P., Bonifacio,
P., Centuri{\'o}n, M., et al. \ 2000, \apj, 541, 54

\bibitem[2001]{nakamura01}
Nakamura, T., Umeda, H., Iwamoto, K.I., et al. 2001, ApJ, 555, 880

\bibitem[1994]{nissen1994}
Nissen, P.E. 1994, Rev. Mex. Astron. Astrofis. 29, 129

\bibitem[2004]{nissen04}
Nissen, P.E., Chen, Y.Q., Asplund, M., \& Pettini, M. 2004, A\&A, 415, 993
(Paper\,I)

\bibitem[1994]{nissen94}
Nissen, P.E., Gustafsson, B., Edvardsson, B., \& Gilmore, G. 1994,
A\&A, 285, 440 

\bibitem[2002]{nissen02}
Nissen, P.E., Primas, F., Asplund, M., \& Lambert, D.L. 2002, A\&A, 390, 235

\bibitem[1997]{nissen97}
Nissen, P.E., \& Schuster, W.J. 1997, A\&A, 326, 751

\bibitem[1997]{nomoto97}
Nomoto, K., Hashimoto, M., Tsujimoto, T. et al. 1997, Nucl. Phys. A, 616, 79

\bibitem[2006]{nomoto06}
Nomoto, K., Tominaga, N., Umeda, H., Kobayashi, C. \& Maeda K. 2006, Nucl. Phys. A, 777, 424

\bibitem[2001]{norris01}
Norris, J.E, Ryan, S.G., \& Beers T.C. 2001, ApJ, 561, 1034

\bibitem[1991]{obrian91}
O'Brian, T.R., Wickliffe, M.E., Lawler, J.E., Whaling, W., \&
Brault, J.W. 1991, J. Opt. Soc. Am. B, 8, 1185

\bibitem[2006]{ohkubo06}
Ohkubo, T., Umeda, H., Maeda, K., et al. 2006, ApJ, 645, 1352

\bibitem[2002]{petitjean02} Petitjean, P.,
Srianand, R., \& Ledoux, C.\ 2002, \mnras, 332, 383

\bibitem[1997]{pettini97}
Pettini, M., King, D.~L., Smith, L.~J., \& Hunstead, R.~W.\ 1997, ApJ, 478, 536

\bibitem[2000]{primas00}
Primas, F., Brugamyer, E., Sneden, C., et al. 2000, in: The First Stars, eds A. Weiss, T. Abel
and V. Hill, ESO Astrophysics Ser., p. 51

\bibitem[2003]{prochaska03} Prochaska, J.~X.,
Howk, J.~C., \& Wolfe, A.~M.\ 2003, \nat, 423, 57

\bibitem[1999]{prochaska99} Prochaska, J.~X.,
\& Wolfe, A.~M.\ 1999, \apjs, 121, 369
 
\bibitem[2002]{prochaska02}
Prochaska, J.~X., \& Wolfe, A.~M.\ 2002, \apj, 566, 68

\bibitem[2007]{prochaska07}
Prochaska, J.~X., Wolfe, A.~M., Howk, J.~C., et al. 2007,
ApJS, in press (astro-ph/0702325)

\bibitem[2004]{przybilla04}
Przybilla, N., \& Butler, K. 2004, ApJ, 610, L61

\bibitem[2000]{ramaty00}
Ramaty, R., Scully, S.T., Lingenfelter, R.E., \& Kozlovsky, B. 2000,
ApJ, 534, 747

\bibitem[2006]{ramirez06}
Ram\'{\i}rez, I., Allende Prieto, C., Redfield, S., \& Lambert, D.L. 
2006, A\&A, 459, 613

\bibitem[2005a]{ramirez05a}
Ram\'{\i}rez, I., \& Melend\'{e}z, J. 2005a, ApJ, 626, 446

\bibitem[2005b]{ramirez05b}
Ram\'{\i}rez, I., \& Melend\'{e}z, J. 2005b, ApJ, 626, 465

\bibitem[2007]{rix07}
Rix, S.~A., Pettini, M., Steidel, C.~C., et al. 2007,
ApJ, submitted

\bibitem[1989]{ryan89}
Ryan, S.G. 1989, AJ, 98, 1693

\bibitem[1999]{ryan99}
Ryan, S.G., Norris, J.E., \& Beers, T.C. 1999, ApJ, 523, 654

\bibitem[2006]{ryde06}
Ryde, N. 2006, A\&A, 455, L13 

\bibitem[2004]{ryde04}
Ryde, N., \& Lambert, D.L. 2004, A\&A, 415, 559

\bibitem[1979]{savage79}
Savage, B.D., \& Mathis, J.S. 1979, ARA\&A, 17, 73


\bibitem[2006]{schuster06}
Schuster, W.J., Moitinho, A., M\'{a}rquez, A., Parrao, L., \&
Covarrubias, E. 2006, A\&A, 445, 939

\bibitem[1988]{schuster88}
Schuster, W.J., \& Nissen, P.E. 1988, A\&AS, 73, 225

\bibitem[1989]{schuster89}
Schuster, W.J., \& Nissen, P.E. 1989, A\&A, 221, 65

\bibitem[1996]{schuster96}
Schuster, W.J., Nissen, P.E., Parrao, L., Beers, T.C., \& Overgaard, L.P.
1996, A\&AS, 117, 317

\bibitem[1993]{schuster93}
Schuster, W.J., Parrao, L., \& Contreras Mart\'{\i}nes, M.E. 
1993, A\&AS, 97, 951

\bibitem[2004]{schnabel04}
Schnabel, R., Schultz-Johanning, M., \& Kock, M. 2004, A\&A, 414, 1169

\bibitem[2006]{skrutskie06}
Skrutskie, M.F., Cutri, R.M., Stiening, R., et al. 2006, AJ, 131, 1163

\bibitem[1999]{stehle99}
Stehl\'{e}, C., \& Hutcheon, R. 1999, A\&AS, 140, 93

\bibitem[1984]{steenbock84}
Steenbock, W., Holweger, H. 1984, A\&A, 130, 319

\bibitem[2002]{takada02}
Takada-Hidai, M., Takeda, Y., Sato, S., et al. 2002, ApJ, 573, 614

\bibitem[2005]{takada05}
Takada-Hidai, M., Saito, Y., Takeda, Y., et al. 2005, PASJ, 57, 347

\bibitem[2005]{takeda05}
Takeda, Y., Hashimoto, O., Taguchi, H., et al. 2005, PASJ, 57, 751

\bibitem[1996]{thielemann96}
Thielemann, F.\,-K., Nomoto, K., \& Hashimoto, M. 1996, ApJ, 460, 408

\bibitem[2000]{vandenberg00}
VandenBerg, D.A., Swenson, F.J., Rogers, F.J., Iglesias, C.A., \& 
Alexander, D.R. 2000, ApJ, 532, 430

\bibitem[2003]{umeda03}
Umeda, H., \& Nomoto, K. 2003, Nature, 422, 871

\bibitem[2005]{umeda05}
Umeda, H., \& Nomoto, K. 2005, ApJ, 619, 427   

\bibitem[2001]{vladilo01}
Vladilo, G., Centuri{\' o}n, M., Bonifacio, P., \& Howk, J.~C.\ 2001,
ApJ, 557, 1007

\bibitem[2005]{wolfe05}
Wolfe, A.M., Gawiser, E., \& Prochaska, J.X. 2005, ARA\&A, 43, 861

\bibitem[1995]{woosley95}
Woosley, S.E., \& Weaver, T.A. 1995, ApJS, 101, 181
\end{thebibliography}
\end{document}